\documentstyle{mn}

\newif\ifAMStwofonts
\AMStwofontstrue

\newcommand{\kev}{keV}
\newcommand{\ergcms}{erg~cm$^{-2}$~s$^{-1}$}

\newcommand{\fe}{Fe~K$\alpha$}
\newcommand{\etal}{et al.}

\ifoldfss
  \ifCUPmtlplainloaded \else
    \NewTextAlphabet{textbfit} {cmbxti10} {}
    \NewTextAlphabet{textbfss} {cmssbx10} {}
    \NewMathAlphabet{mathbfit} {cmbxti10} {} 
    \NewMathAlphabet{mathbfss} {cmssbx10} {} 
  \fi
  \ifAMStwofonts
    \ifCUPmtlplainloaded \else
      \NewSymbolFont{upmath} {eurm10}
      \NewSymbolFont{AMSa} {msam10}
      \NewMathSymbol{\upi}     {0}{upmath}{19}
      \NewMathSymbol{\umu}     {0}{upmath}{16}
      \NewMathSymbol{\upartial}{0}{upmath}{40}
      \NewMathSymbol{\leqslant}{3}{AMSa}{36}
      \NewMathSymbol{\geqslant}{3}{AMSa}{3E}

    \fi
  \fi
\fi 

\ifnfssone
  \newmathalphabet{\mathit}
  \addtoversion{normal}{\mathit}{cmr}{m}{it}
  \addtoversion{bold}{\mathit}{cmr}{bx}{it}
  \newmathalphabet{\mathbfit} 
  \addtoversion{normal}{\mathbfit}{cmr}{bx}{it}
  \addtoversion{bold}{\mathbfit}{cmr}{bx}{it}
  \newmathalphabet{\mathbfss} 
  \addtoversion{normal}{\mathbfss}{cmss}{bx}{n}
  \addtoversion{bold}{\mathbfss}{cmss}{bx}{n}
  \ifAMStwofonts
    \ifCUPmtlplainloaded \else
      \UseAMStwoboldmath
      \makeatletter
      \new@mathgroup\upmath@group
      \define@mathgroup\mv@normal\upmath@group{eur}{m}{n}
      \define@mathgroup\mv@bold\upmath@group{eur}{b}{n}
      \edef\UPM{\hexnumber\upmath@group}
      \new@mathgroup\amsa@group
      \define@mathgroup\mv@normal\amsa@group{msa}{m}{n}
      \define@mathgroup\mv@bold\amsa@group{msa}{m}{n}
      \edef\AMSa{\hexnumber\amsa@group}
      \makeatother
      \mathchardef\upi="0\UPM19
      \mathchardef\umu="0\UPM16
      \mathchardef\upartial="0\UPM40
      \mathchardef\leqslant="3\AMSa36
      \mathchardef\geqslant="3\AMSa3E
    \fi
  \fi
\fi 

\ifnfsstwo
  \DeclareMathAlphabet{\mathbfit}{OT1}{cmr}{bx}{it}
  \SetMathAlphabet\mathbfit{bold}{OT1}{cmr}{bx}{it}
  \DeclareMathAlphabet{\mathbfss}{OT1}{cmss}{bx}{n}
  \SetMathAlphabet\mathbfss{bold}{OT1}{cmss}{bx}{n}
  \ifAMStwofonts
    \ifCUPmtlplainloaded \else
      \DeclareSymbolFont{UPM}{U}{eur}{m}{n}
      \SetSymbolFont{UPM}{bold}{U}{eur}{b}{n}
      \DeclareSymbolFont{AMSa}{U}{msa}{m}{n}
      \DeclareMathSymbol{\upi}{0}{UPM}{"19}
      \DeclareMathSymbol{\umu}{0}{UPM}{"16}
      \DeclareMathSymbol{\upartial}{0}{UPM}{"40}
      \DeclareMathSymbol{\leqslant}{3}{AMSa}{"36}
      \DeclareMathSymbol{\geqslant}{3}{AMSa}{"3E}
    \fi
  \fi
\fi 

\ifCUPmtlplainloaded \else
  \ifAMStwofonts \else 
    \def\upi{\pi}
    \def\umu{\mu}
    \def\upartial{\partial}
  \fi
\fi

\title[X-ray reflection by photoionized accretion discs]
  {X-ray reflection by photoionized accretion discs}
\author[D.\ R.\ Ballantyne, R.\ R.\ Ross \& A.\ C.\ Fabian]
  {D.~R.~Ballantyne$^{1}$\thanks{drb@ast.cam.ac.uk}, 
  R.~R.~Ross$^{1,2}$ and A.~C.~Fabian$^1$\\
  $^1$Institute of Astronomy, Madingley Road, Cambridge CB3 0HA \\
  $^2$Physics Department, College of the Holy Cross, Worcester, MA 01610, USA}
\pagerange{\pageref{firstpage}--\pageref{lastpage}}
\pubyear{2001}

\input{psfig.sty}

\begin{document}

\label{firstpage}

\maketitle

\begin{abstract}
Calculations of X-ray reflection spectra from ionized, optically-thick
material are an important tool in investigations of accretion flows
surrounding compact objects. We present the results of reflection
calculations that treat the relevant physics with a minimum of
assumptions. The temperature and ionization structure of the top five
Thomson depths of an illuminated disc are calculated while also
demanding that the atmosphere is in hydrostatic equilibrium. In
agreement with Nayakshin, Kazanas \& Kallman, we find that there is a
rapid transition from hot to cold material in the illuminated
layer. However, the transition is usually not sharp so that often we find a
small but finite region in Thomson depth where there is a stable
temperature zone at $T \sim 2 \times 10^{6}$~K due to photoelectric
heating from recombining ions. As a result, the reflection spectra
often exhibit strong features from partially-ionized material,
including helium-like Fe~K lines and edges.  The reflection spectra,
when added to the illuminating spectra, were  fit by the publicly available constant density models (i.e., {\sc pexriv}, {\sc pexrav} and the models of Ross \& Fabian). We find that due to the highly ionized features in the spectra these models have difficulty correctly parameterizing the new reflection spectra. There is evidence for a spurious $R-\Gamma$ correlation in the
\textit{ASCA} energy range, where $R$ is the reflection fraction for a
power-law continuum of index $\Gamma$, confirming the suggestion of
Done \& Nayakshin that at least part of the $R-\Gamma$ correlation
reported by Zdziarski, Lubi\'{n}ski \& Smith for Seyfert galaxies and
X-ray binaries might be due to ionization effects. However, large
uncertainties in the fit parameters prevent confirmation of the
correlation in the 3--20~\kev\ energy range. Although many of
the reflection spectra show strong ionized features, these are not
typically observed in most Seyfert and quasar X-ray spectra. However,
the data are not yet good enough to place constraints on the
illumination properties of discs, as instrumental and/or relativistic
effects could mask the ionized features predicted by the models.
\end{abstract}


\section{Introduction}
\label{sect:intro}

The X-ray spectra from luminous accreting black holes, both of stellar
mass and in the nuclei of galaxies, generally consist of a thermal
component attributed to viscous dissipation in a thin dense disc and a
harder power-law component attributed to an optically-thin corona
above the disc. Irradiation of the thin disc by the corona yields
another component, termed the reflection spectrum. The spectral shape
of this component depends in detail upon the incident coronal emission
as well as the ionization state of the surface layers of the disc.
Hydrogen and helium are of course highly ionized, but heavier elements
such as oxygen, silicon and particularly iron may not be completely
stripped of electrons and so can impart characteristic spectral
features, lines and edges, into the observed spectrum. Such features
can be of great importance since they carry information about the disc
material, such as its ionization state and composition, as well as its
motion and depth in the gravitational well of the black hole.

Initial studies of such photoionized discs have commonly assumed that
the surface density is constant with depth (Ross \& Fabian 1993; Matt,
Fabian \& Ross 1993a,b; \.{Z}ycki \etal\ 1994; Ross, Fabian \& Young
1999).  Constant density may be an appropriate approximation for the
bulk of a radiation-pressure-supported disc. Recently, Nayakshin and
colleagues (Nayakshin, Kazanas \& Kallman 2000; Nayakshin \& Kallman
2001; Done \& Nayakshin 2001; Nayakshin 2000) have studied the
situation in which the number density decreases with height in accord
with hydrostatic equilibrium. Their results often show significant
differences with those from the constant density calculations: the
reflection spectrum in the 2--20~keV band consists mostly of a
featureless spectrum from reflection by highly ionized outer gas, and
a weak low ionization component from deeper recombined gas. This major
difference is attributed to a thermal instability (e.g., Krolik, McKee
\& Tarter 1981) that can cause a sharp transition in the density and
temperature profiles. However, Nayakshin \etal\ (2000) artificially
vary the relative strength of gravity on the atmosphere (parameterized
by their $A$ parameter; eq.~\ref{eqn:A}) in an attempt to account for
any effects by a local outflow due to heating by magnetic flares above
the disc.

Here we report on our study of the reflection spectra from disc
atmospheres in hydrostatic equilibrium, using the code of Ross (1978).
This is a modified version of the one with which we studied constant
density atmospheres (Ross \& Fabian 1993). We take a complementary
approach to Nayakshin \etal\ (2000), and investigate the response of
the reflection spectrum to changes in the disc and illumination
parameters. These models will be more appropriate to fit to
forthcoming {\it XMM-Newton} and {\it Chandra} data. We find that the gas
temperature does drop rapidly at some point within the surface, but
generally the outer, more highly ionized gas just above that point
emits strong lines from hydrogenic and helium-like iron.  Reflection
from the outer, most highly ionized gas does add a featureless
component to the spectrum. Our results approximately resemble a dilute
version of the constant density spectra.

The paper is organized as follows. In Section~\ref{sect:comp}, we
outline the changes we made to the Ross \& Fabian (1993) code to
include hydrostatic equilibrium. The following section
(\S~\ref{sect:res}) presents the results of the calculations.  In
Section~\ref{sect:compare} we illustrate how the new calculations
differ from earlier constant density ones.  We fit simulated spectra
comprised from the new models with the older constant density models
in Section~\ref{sect:const-dens}.  Section~\ref{sect:limits} contains
a discussion on the limitations of the code as it now stands. We
discuss the implications of our results in Sect.~\ref{sect:discuss},
and summarize the conclusions in Section~\ref{sect:concl}.

\section{Computations}
\label{sect:comp}
We model the uppermost layer of an accretion disc that is illuminated
directly from above, as by local magnetic flares.  This atmosphere is
treated down to a Thomson depth $\tau_{\rm T}=5$, so that little of
the X-ray illumination reaches the bottom of this layer without first
interacting with the gas.  The external illumination is assumed to
have the form of a power-law spectrum with photon index $\Gamma$
(i.e., photon flux $\propto E^{-\Gamma}$) and extends from 1~eV to a
sharp cutoff at 100~\kev.  The radiation entering the atmosphere from
the disc below is in the form of an effective blackbody spectrum with
total flux given by the Newtonian value \cite{sha73},
\begin{equation}
F_{\rm disc}=\frac{3GM\dot M}{8\pi R^3}
 \left[1-\left(\frac{3R_{\rm S}}{R}\right)^{1/2}\right],
\label{eq:softflux}
\end{equation}
where $M$ is the mass of the central black hole, $\dot M$ is the accretion 
rate, $R$ is the radial distance away from the black hole, and $R_{\rm S}$ 
is the Schwarzschild radius.  (For an accretion disc extending 
down to $3R_{\rm S}$, the resulting efficiency is 
$\eta\equiv L/\dot Mc^2=1/12$.) The base of the atmosphere is fixed at a 
vertical height above the midplane of the disc equal to the expected 
half-thickness of a gas-pressure dominated disc with $\alpha=0.1$ \cite{mer00},
\begin{equation}
H=(8.33\times 10^3{\rm\thinspace cm})\dot m^{1/5}m^{9/10}r^{21/20}
 \left(1-\sqrt{\frac{3}{r}}\right)^{1/5},
\label{eq:height}
\end{equation}
where $m=M/M_{\odot}$, $r=R/R_{\rm S}$, and 
$\dot m=\dot M/\dot M_{\rm Edd}$ 
($\dot M_{\rm Edd}=4 \pi G m_{p} M/ \eta \sigma_{\rm T} c$, where $m_p$ is 
the proton mass, is the Eddington accretion rate).

The method of treating the radiative transfer has been described by
Ross \& Fabian (1993).  The penetration of X-rays from the external
illumination is treated analytically in a one-stream approximation.
The diffuse radiation in the atmosphere resulting from the impingement
of the disc flux, Compton scattering of the illuminating radiation,
and emission within the gas is treated using the
Fokker-Planck/diffusion method of Ross, Weaver \& McCray
\shortcite{ros78} in plane-parallel geometry.  The local temperature
and ionization state of the gas is found by solving the equations of
thermal and ionization equilibrium, so that they are consistent with
the local radiation field.  Hydrogen and helium are assumed to be
fully ionized, while the following ionization stages of the most
abundant metals are treated: C~{\sc v--vii}, O~{\sc v--ix}, Mg~{\sc
ix--xiii}, Si~{\sc xi--xv}, and Fe~{\sc xvi--xxvii}.

The density of the gas is found from the condition for hydrostatic 
equilibrium.  Let us denote the net upward spectral flux at height $z$ 
above the midplane of the disc by
\begin{equation}
{\cal F}_{\nu}(z)=F_{\nu}(z)-X_{\nu}(z),
\end{equation}
where $F_{\nu}$ is the contribution due to diffuse radiation and 
$X_{\nu}$ is the magnitude of the contribution due to direct inward 
penetration of the external illumination.  In the thin-disc approximation, 
the equation of hydrostatic equilibrium in the vertical direction is
\begin{equation}
-\frac{dP_{\rm gas}}{dz}
 +\int_0^{\infty}\rho\kappa_{\nu}\frac{F_{\nu}}{c}d\nu 
 -\int_0^{\infty}\rho\kappa_{\nu}\frac{X_{\nu}}{c}d\nu
 =\frac{GM\rho}{R^3}z,
\label{eq:hetop} 
\end{equation}
where $P_{\rm gas}$ is the gas pressure, $\rho$ is the gas density, and
$\kappa_{\nu}$ is the opacity.  At interior points, the diffuse flux is 
given by
\begin{equation}
F_{\nu}=-\frac{c}{3\rho\kappa_{\nu}}\frac{du_{\nu}}{dz},
\end{equation}
where $u_{\nu}$ is the spectral energy density.  Therefore, except at the
boundaries of the atmosphere, equation~(\ref{eq:hetop}) simplifies to
\begin{equation}
-\frac{dP_{\rm gas}}{dz} -\frac{1}{3}\frac{du}{dz} 
 -\int_0^{\infty}\rho\kappa_{\nu}\frac{X_{\nu}}{c}d\nu
 =\frac{GM\rho}{R^3}z,
\label{eq:hez} 
\end{equation}
where $u$ is the total energy density in diffuse radiation.  Dividing 
through by $n_{\rm e}\sigma_{\rm T}$, where $n_{\rm e}$ is the number 
density of free electrons and $\sigma_{\rm T}$ is the Thomson cross 
section, equation~(\ref{eq:hez}) gives
\begin{equation}
\frac{dP_{\rm gas}}{d\tau_{\rm T}} 
 +\frac{1}{3}\frac{du}{d\tau_{\rm T}} 
 -\int_0^{\infty}\frac{\rho\kappa_{\nu}}{n_{\rm e}\sigma_{\rm T}}
 \frac{X_{\nu}}{c}d\nu
 =\frac{GM\rho}{n_{\rm e}\sigma_{\rm T}R^3}z,
\label{eq:hetau}  
\end{equation}
where $\tau_{\rm T}$ is the Thomson depth inward from the surface.
With helium at one-tenth the number abundance of hydrogen, the relations
$P_{\rm gas}=2.3n_{\rm H}kT$, $\rho=1.4n_{\rm H}m_{\rm H}$, and
$n_{\rm e}=1.2n_{\rm H}$ are used to make the dependent variable
$n_{\rm H}$, the number density of hydrogen.

We use the following procedure to model the illuminated atmosphere.  The
solution is sought over a grid of 110 zones in Thomson depth $\tau_{\rm T}$ 
by 210 bins in photon energy $E$.  The initial density distribution is 
taken to be a Gaussian with scale height equal to the half-thickness of 
the disc below,
\begin{equation}
n_{\rm H}(z)=n_0\exp\left[-\frac{(z-H)^2}{H^2}\right],
\end{equation}
with base density $n_0$ chosen so that the atmosphere has the 
desired total Thomson depth.  With the density structure fixed, the
temperature structure and radiation field throughout the gas are
determined as described by Ross \& Fabian (1993).  

This temperature structure and radiation
field are then used to find an implied density structure.  
Starting at the current estimate for the top height of the atmosphere 
and assuming that the gas pressure is negligibly small above it, 
equation~(\ref{eq:hetop}) is used to find the value of $n_{\rm H}$ in
the uppermost zone.  This density and the desired value of $\tau_{\rm T}$
at the base of the uppermost zone then establishes the height ($z$) there.
Working inward through the atmosphere, equation~(\ref{eq:hetau}) is used
to find the value of $n_{\rm H}$ and the base height for subsequent zones.
When the bottom zone is reached, the height implied for the base of the 
atmosphere, $z_{\rm base}$, is compared with the desired height $H$.  The
estimate for the top height of the atmosphere is adjusted and the 
integration inward is repeated until the proper base height,
$z_{\rm base}=H$, is achieved.

With a new estimation of the density structure established, the 
temperature structure and radiation field throughout the gas are
recalculated.  The procedures for finding the density structure, the
temperature structure, and the radiation field are repeated until the
model converges. Our convergence condition was that all of the
following must be true: a) at least 15 iterations must have been
completed (this guarantees that the various ionization and recombination
fronts have had time to propagate through the model), b) the fractional change
in the top height from the previous iteration is less than 10$^{-4}$, 
c) the fractional change in the surface temperature is less than 10$^{-3}$, 
and d) the fractional change in the surface density is less than 10$^{-3}$.
In almost all cases, the models did not require many more iterations than the 
15 they were originally allocated.

\section{Results}
\label{sect:res}

Our models require six input parameters: three describing properties
of the disc, and three describing properties of the illuminating
radiation. The three disc properties are the black
hole mass $m$, the accretion rate $\dot m$, and the radius along the
accretion disc $r$.  The three parameters
that describe the incident radiation are the photon-index of the
illuminating power-law $\Gamma$, the flux $F_x$ (in \ergcms), and the
incidence angle $i$ (measured from the normal). In the following
subsections we will show the results of the models varying each of the
parameters in turn, deferring a detailed quantitative discussion of
the results to later sections.  Our canonical model, from
which all variations were originated, had the following
parameters: $m=10^{8}$, $\dot m=0.001$, $r=9$, $\Gamma=1.9$,
$F_x=10^{15}$, and $i=\cos^{-1}(1/\sqrt{3})=54\fdg7$.

To facilitate comparison with the results of Nayakshin \etal\ (2000), we
also calculate their parameter $A$, where
\begin{equation}
A={GM\rho H c \over R^3 \sigma_{\rm T} n_{\rm e} F_x}
={5.42 \times 10^{20} H \over m^2 r^3 F_x}.
\label{eqn:A}
\end{equation}
For our canonical model, $A=0.0207$. We have summarized all the models
presented in this section, along with their $A$ values, in
Table~\ref{table:models}.
\begin{table*}
\begin{minipage}{110mm}
\caption{A summary of the parameters used to calculate the models
presented in Section~\ref{sect:res}.}
\label{table:models}
\begin{tabular}{@{}lcccccccc}
Type of & $\Gamma$ & $F_x$ & $i$ & $r$ & $m$ & $\dot m$ & $A$ & $F_x / F_{\rm disc}$ \\
Model & & (\ergcms) & (degrees) & & & & & \\ \hline\hline
Canonical & 1.9 & 10$^{15}$ & 54.7 & 9 & 10$^8$ & 0.001 & 0.0207 & 144\\ \hline
Vary $\Gamma$ & 2.2 & -- & -- & -- & -- & -- & -- & --\\
 & 2.1 & -- & -- & -- & -- & -- & -- & -- \\
 & 2.0 & -- & -- & -- & -- & -- & -- & -- \\
 & 1.8 & -- & -- & -- & -- & -- & -- & -- \\
 & 1.7 & -- & -- & -- & -- & -- & -- & -- \\
 & 1.6 & -- & -- & -- & -- & -- & -- & -- \\ \hline
Vary $F_x$ & 1.9 & 10$^{16\ddag}$ & -- & -- & -- & -- & 0.0021 & 1442 \\
 & -- & $5 \times 10^{15\ddag}$ & -- & -- & -- & -- & 0.0041 & 721 \\
 & -- & $5 \times 10^{14}$ & -- & -- & -- & -- & 0.0414 & 72 \\
 & -- & 10$^{14}$ & -- & -- & -- & -- & 0.2068 & 14 \\ \hline
Vary $i$ & -- & 10$^{15}$ & 20 & -- & -- & -- & 0.0207 & 144 \\
 & -- & -- & 80 & -- & -- & -- & -- & -- \\ \hline
Vary $r$ & -- & -- & 54.7 & 4 & -- & -- & 0.0799 & 40 \\
 & -- & -- & -- & 6 & -- & -- & 0.0424 & 62 \\
 & -- & -- & -- & 8 & -- & -- & 0.0256 & 110 \\
 & -- & -- & -- & 10 & -- & -- & 0.0171 & 185 \\
 & -- & -- & -- & 12 & -- & -- & 0.0122 & 289 \\
 & -- & -- & -- & 16 & -- & -- & 0.0071 & 604 \\
 & -- & -- & -- & 18$^{\ddag}$ & -- & -- & 0.0057 & 824 \\ \hline
Vary $m$ & -- & -- & -- & 9 & 10$^{9\ddag}$ & -- & 0.0016 & 1442 \\
 & -- & -- & -- & -- & 10$^7$ & -- & 0.2604 & 14 \\ \hline
Vary $\dot m$ & -- & -- & -- & -- & 10$^8$ & 0.1$^{\dag}$ & 0.8428 & 1.4 \\
 & -- & -- & -- & -- & -- & 0.05$^{\dag}$ & 0.4214 & 2.9 \\
 & -- & -- & -- & -- & -- & 0.01 & 0.0328 & 14 \\
 & -- & -- & -- & -- & -- & 0.005 & 0.0285 & 29 \\
 & -- & -- & -- & -- & -- & 0.0025 & 0.0248 & 58 \\
 & -- & -- & -- & -- & -- & 0.0005 & 0.0180 & 288 \\ \hline
\end{tabular}
\medskip

$^{\dag}$ utilized radiation pressure boundary conditions

$^{\ddag}$ calculated with a $\tau_{T}=10$ deep atmosphere
\end{minipage}
\end{table*}
\subsection{Varying the photon index}
\label{sub:phindex}

\begin{figure}
\centerline{\psfig{figure=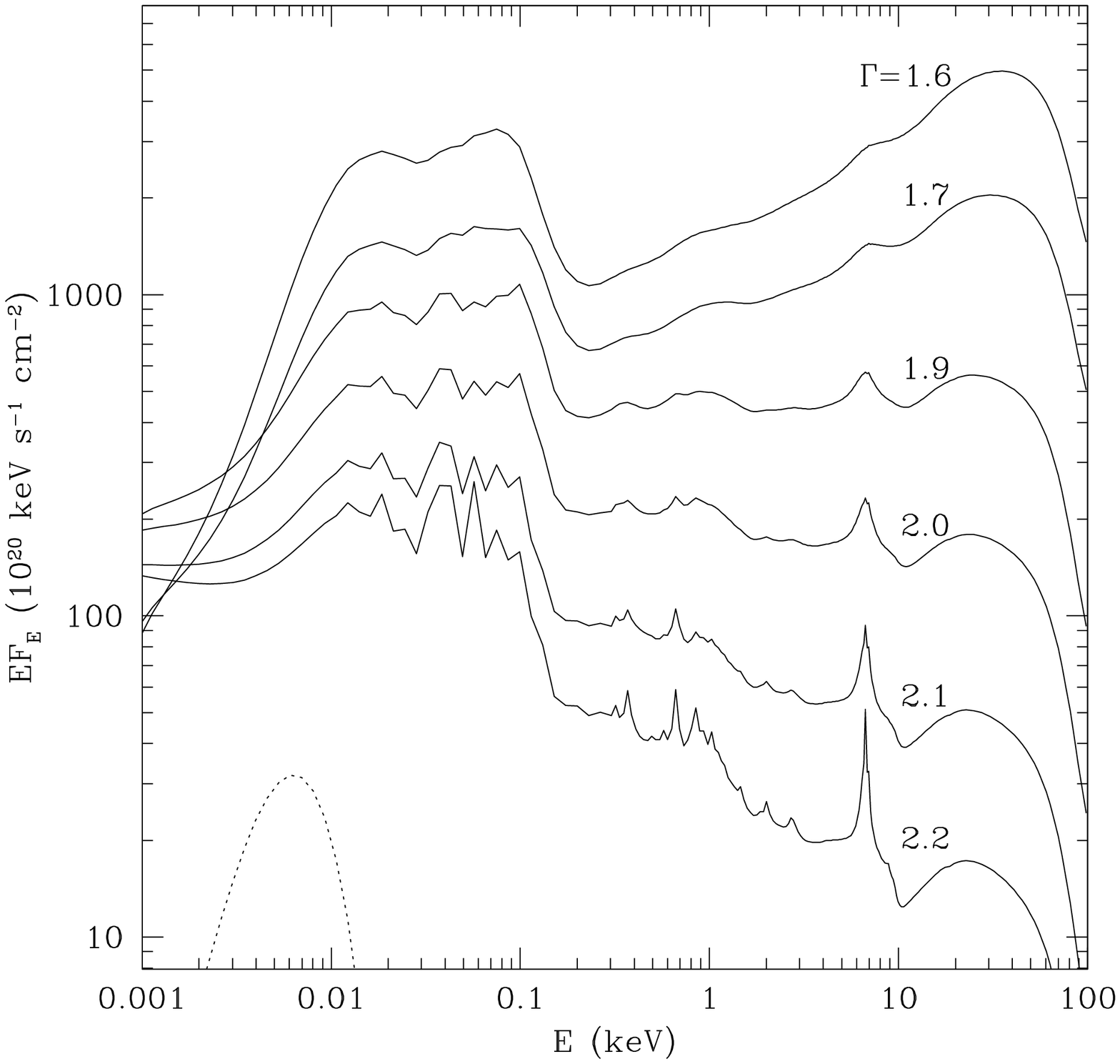,width=0.50\textwidth,silent=}}
\centerline{\psfig{figure=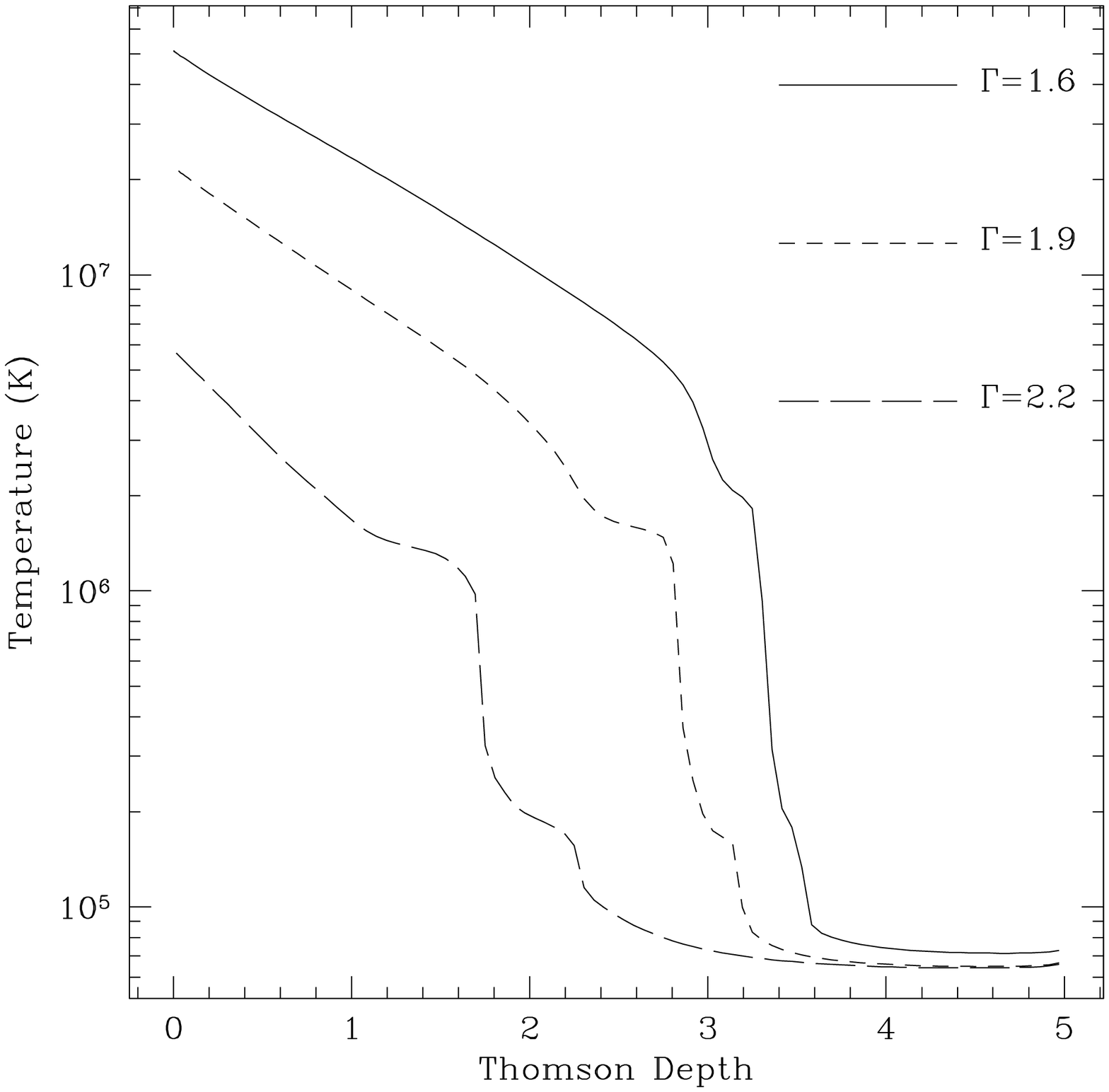,width=0.50\textwidth,silent=}}
\caption{{\it Top}: Reflection spectra for models where the
photon-index of the illuminating power-law has been varied. All the
other parameters have the same values as in the canonical model, so
all these models have the value of $A=0.0207$. The spectra have been
offset vertically for clarity.  The dotted curve shows the spectrum
from the disc that is incident on the base of the atmosphere. {\it
Bottom}: The temperature of the atmosphere as a function of Thomson
depth for three different values of $\Gamma$. The harder the spectrum,
the deeper into the atmosphere it ionized, and the sharper the thermal
transition.}
\label{fig:gamma}
\end{figure}

In the top panel of Figure~\ref{fig:gamma}, we present our calculated
reflection spectra for six values of the photon index $\Gamma$.
The bottom panel shows how the temperature of the atmosphere varies
with Thomson depth for three different values of $\Gamma$. The harder
spectra ionize further into the slab than the softer ones, and
have a sharper thermal transition. This results in a highly ionized
and reflective spectrum as is seen in the top panel in
Figure~\ref{fig:gamma}. The softer illuminating spectra result in a
much more complex reflection spectrum, exhibiting features from a variety of
ionization states.

\begin{figure}
\centerline{\psfig{figure=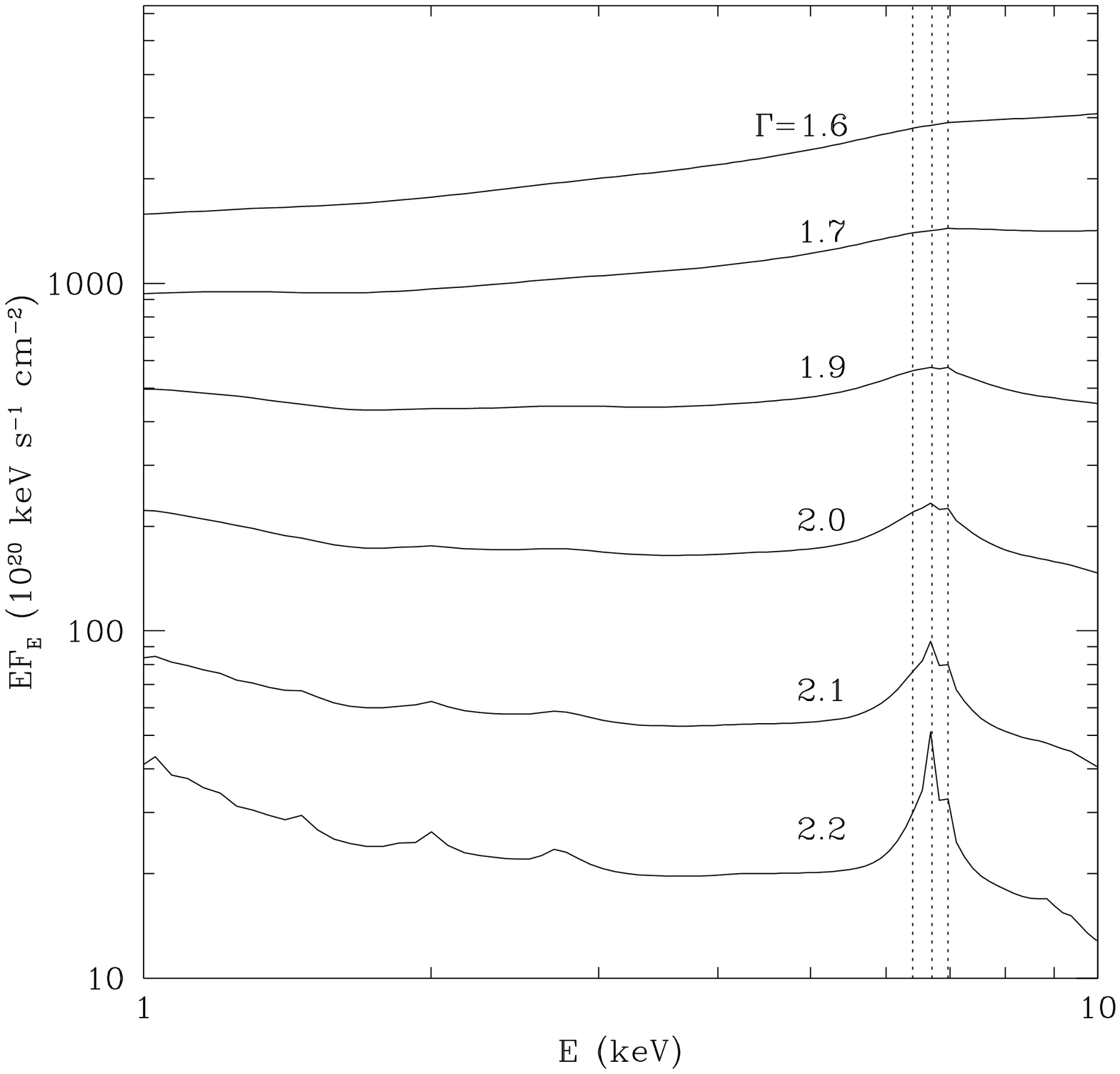,width=0.50\textwidth,silent=}}
\caption{The same reflection spectra as shown in Fig.~\ref{fig:gamma},
but plotted between 1 and 10~\kev\ to emphasize the features of the
\fe\ line. Again, the spectra have been offset vertically for ease of
presentation. The dotted vertical lines show the positions of the iron
lines at 6.4, 6.7 and 7.0~\kev. The ionized Fe lines at 6.7 and
7.0~\kev\ are found in the reflection spectra, irrespective of the
photon-index of the illuminating continuum.} 
\label{fig:gamma-fe}
\end{figure}

To emphasize the properties of the \fe\ line, we present in 
Figure~\ref{fig:gamma-fe} the reflection spectra plotted between 1 and
10~\kev.
The \fe\ line is only barely visible when the illuminating spectrum is very
hard, and then becomes more pronounced as the incident power-law
softens. However, we find that when the disc is highly illuminated (as
it is in all our models) {\it at no point does a neutral Fe line at
6.4~\kev\ become prominent.} The Fe line complex is made up almost
entirely from the ionized lines at 6.7 and 7.0~\kev. 
 
\subsection{Varying the illuminating flux}
\label{sub:flux}

\begin{figure}
\centerline{\psfig{figure=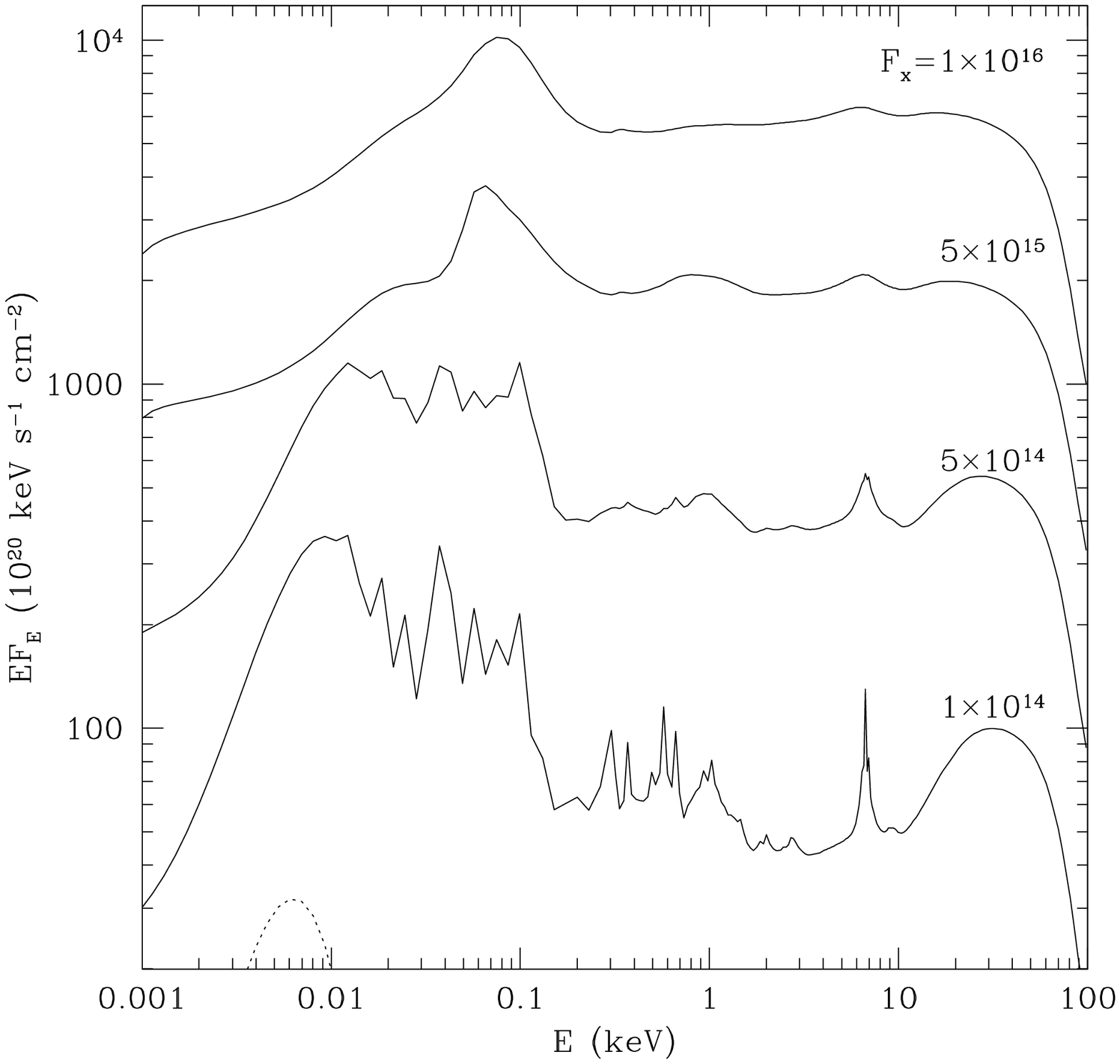,width=0.50\textwidth,silent=}}
\centerline{\psfig{figure=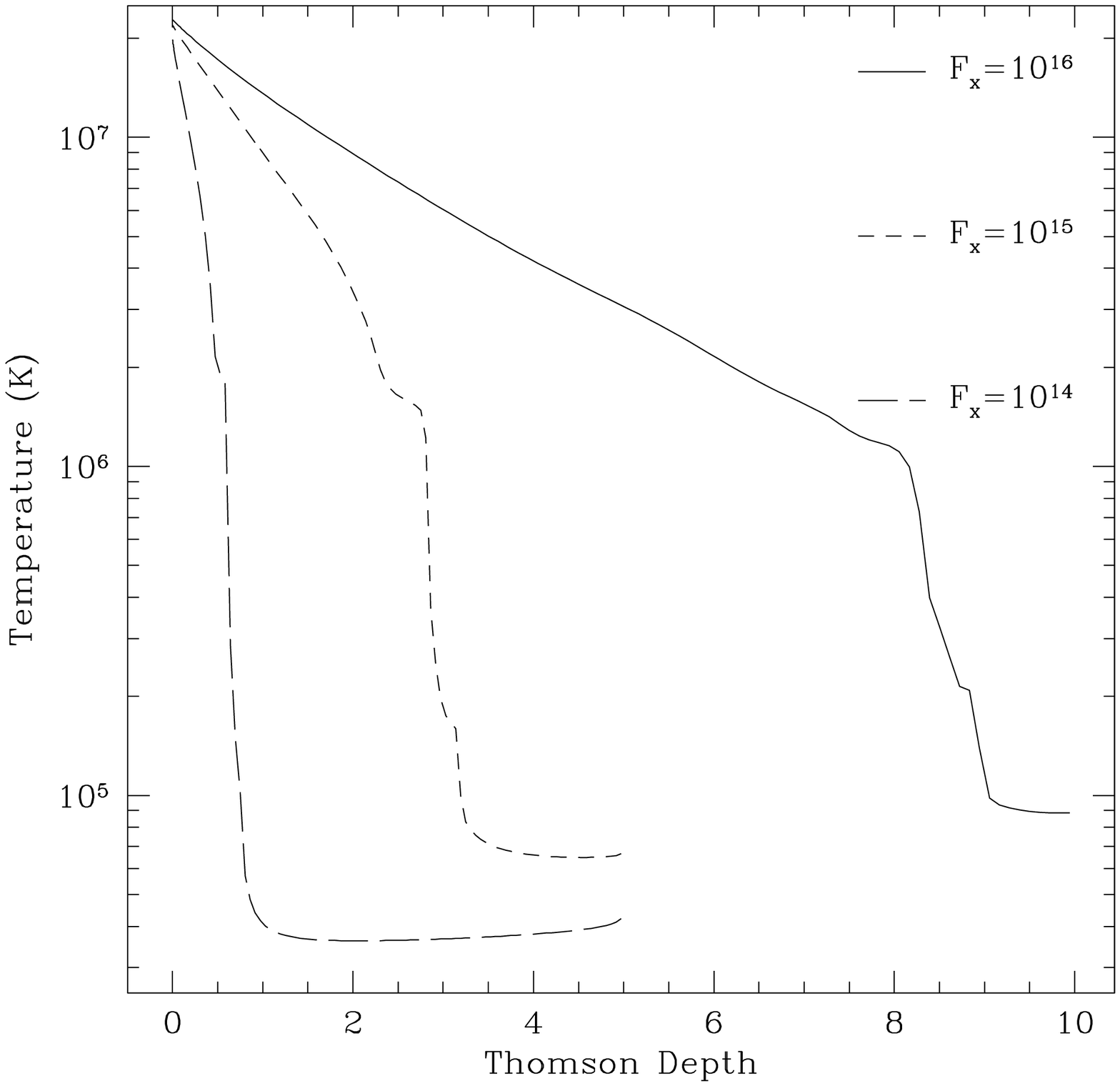,width=0.50\textwidth,silent=}}
\caption{{\it Top}: Reflection spectra for models where the
illuminating flux has been varied. All the other parameters have the
same values as in the canonical model. The spectra have been offset
vertically for clarity. The dotted curve shows the spectrum from the
disc that is incident on the base of the atmosphere. Since the models with $F_x=10^{16}$ and $5 \times 10^{15}$~\ergcms\ ionize through 5 Thomson depths, they were calculated with the atmosphere extending down to $\tau_{\rm T}=10$. The reflection spectrum for the least intense illuminating flux
is the only one with many ionized features. {\it Bottom}:
The temperature of the atmosphere as a function of Thomson depth for
three different values of $F_x$. The least intense radiation field has hardly any mid-temperature shoulder and does not ionize very far into the
atmosphere. In all cases, the units of $F_x$ are \ergcms.}
\label{fig:fx}
\end{figure}

Next, we consider models where the illuminating flux is varied over 
two orders of magnitude. Figure~\ref{fig:fx} shows
the reflection spectra and the temperature structure for some of these
models.
The illuminating radiation fields with $F_x=10^{16}$ and $5 \times 10^{15}$~\ergcms\ ionize through 
five Thomson depths so these models were calculated with the atmosphere extending down to $\tau_{\rm T}=10$. As
the illuminating flux is wound down, the temperature transition
becomes much sharper, and moves closer to the surface of the
atmosphere. Indeed, the reflection spectrum for the model with
$F_x=10^{14}$~\ergcms, is the only one out of the four to show many ionized lines.

\subsection{Varying the incidence angle}
\label{sub:angle}

\begin{figure}
\centerline{\psfig{figure=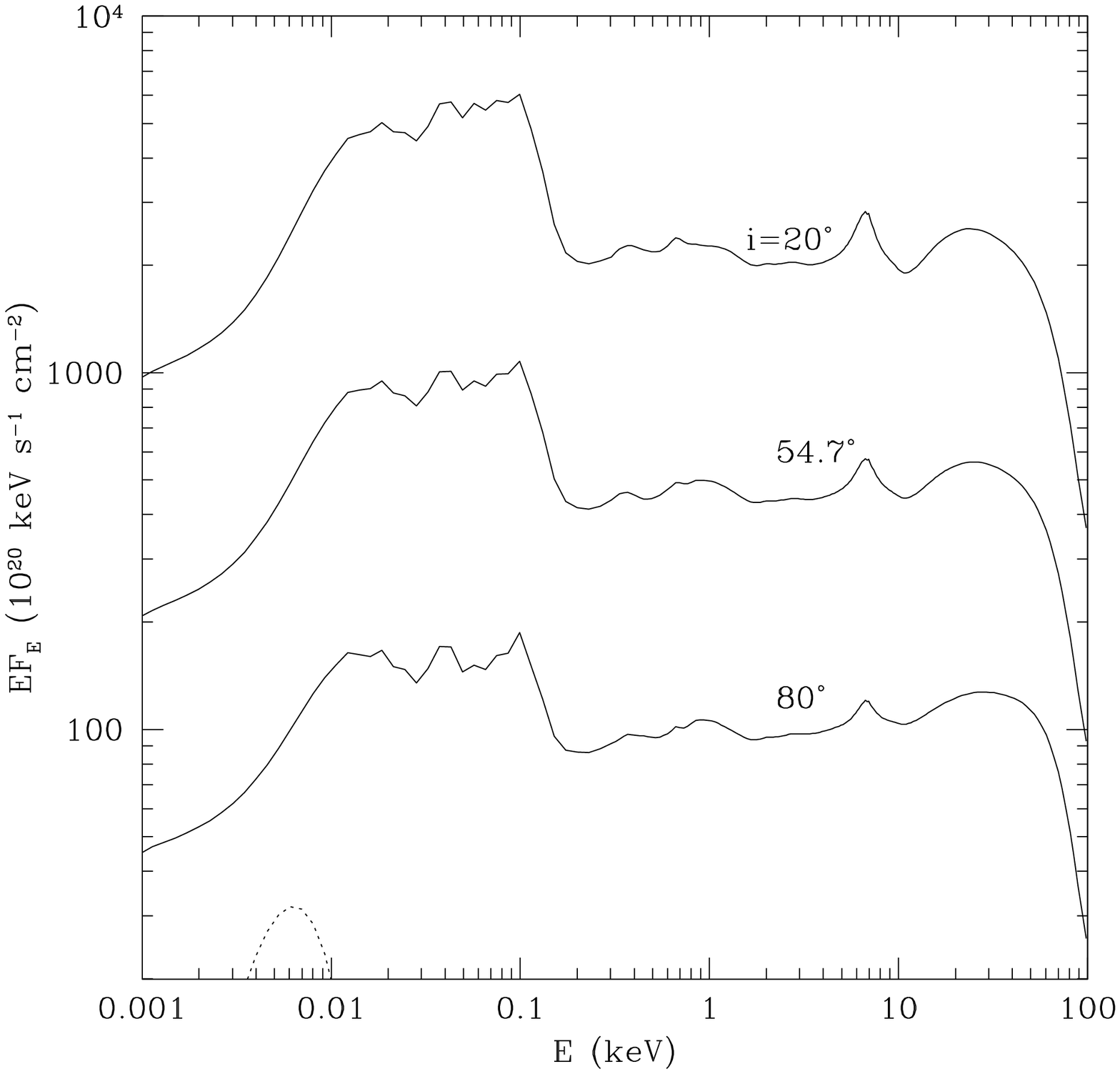,width=0.50\textwidth,silent=}}
\centerline{\psfig{figure=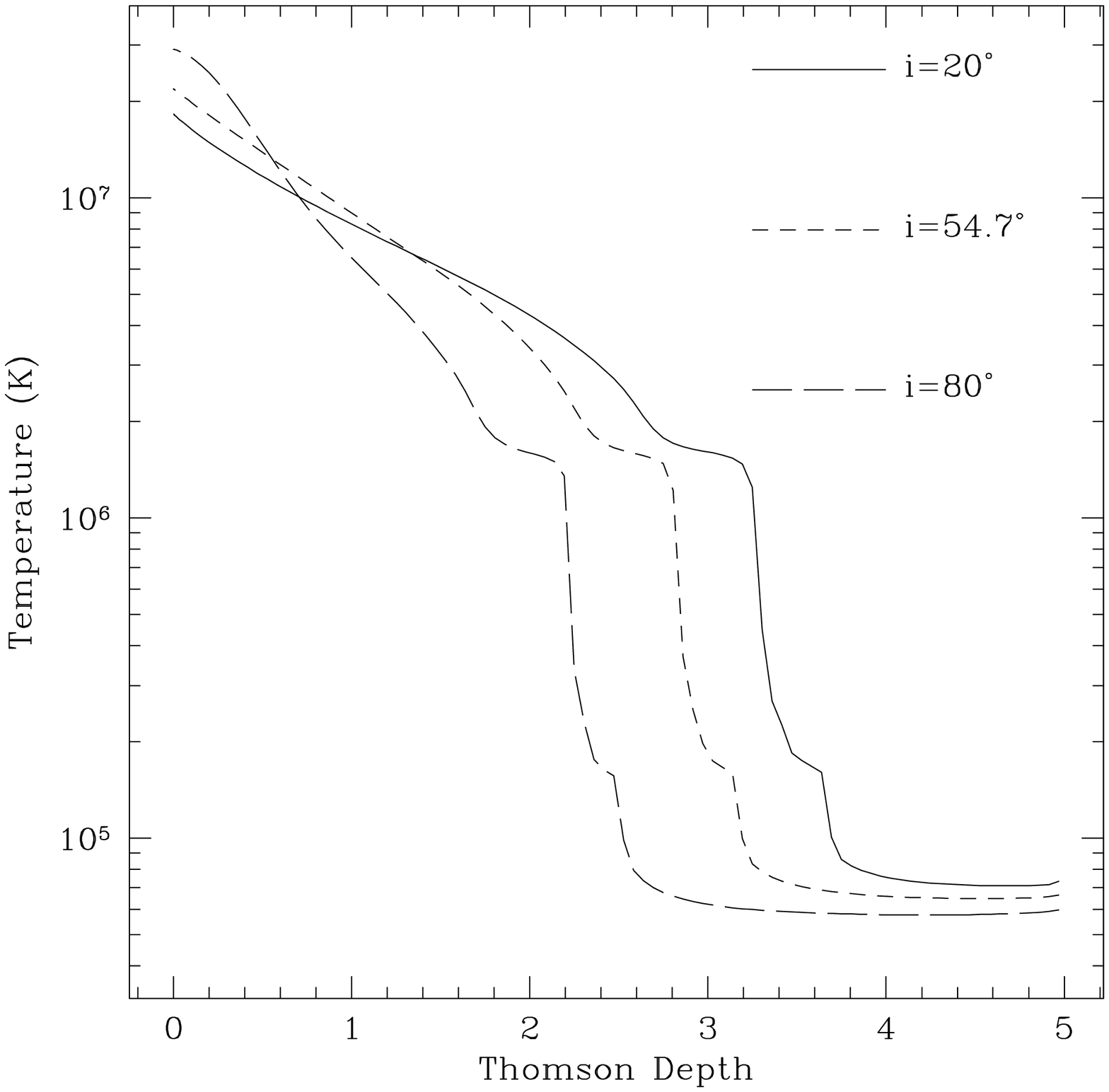,width=0.50\textwidth,silent=}}
\caption{{\it Top}: Reflection spectra for models where the incidence
angle of the radiation has been varied. All the other parameters have
the same values as in the canonical model. The spectra have been
offset vertically for clarity. The dotted curve shows the spectrum
from the disc that is incident on the base of the atmosphere.
{\it Bottom}: The temperature of the atmosphere as a function of
Thomson depth for three different values of $i$.  The most direct
radiation ionizes deeper into the atmosphere.}
\label{fig:angle}
\end{figure}

Our canonical model assumes as a crude approximation to isotropic
radiation that the incident photons are striking the atmosphere at an
incidence angle of $i=54\fdg7$, such that $\mu=\cos
i=1/\sqrt{3}$. Figure~\ref{fig:angle} shows the results when we change
the angle of incidence to 20$^{\circ}$ and to 80$^{\circ}$. The
incidence angle is measured with respect to the normal, so that large
angles correspond to grazing incidence.

The results are not surprising. The radiation that is more directly
illuminating the atmosphere ionizes deeper into the slab than the
radiation that has a grazing incidence. However, at these high flux levels there is little effect on the emergent reflection spectrum.

\subsection{Varying the radius}
\label{sub:radius}
\begin{figure}
\centerline{\psfig{figure=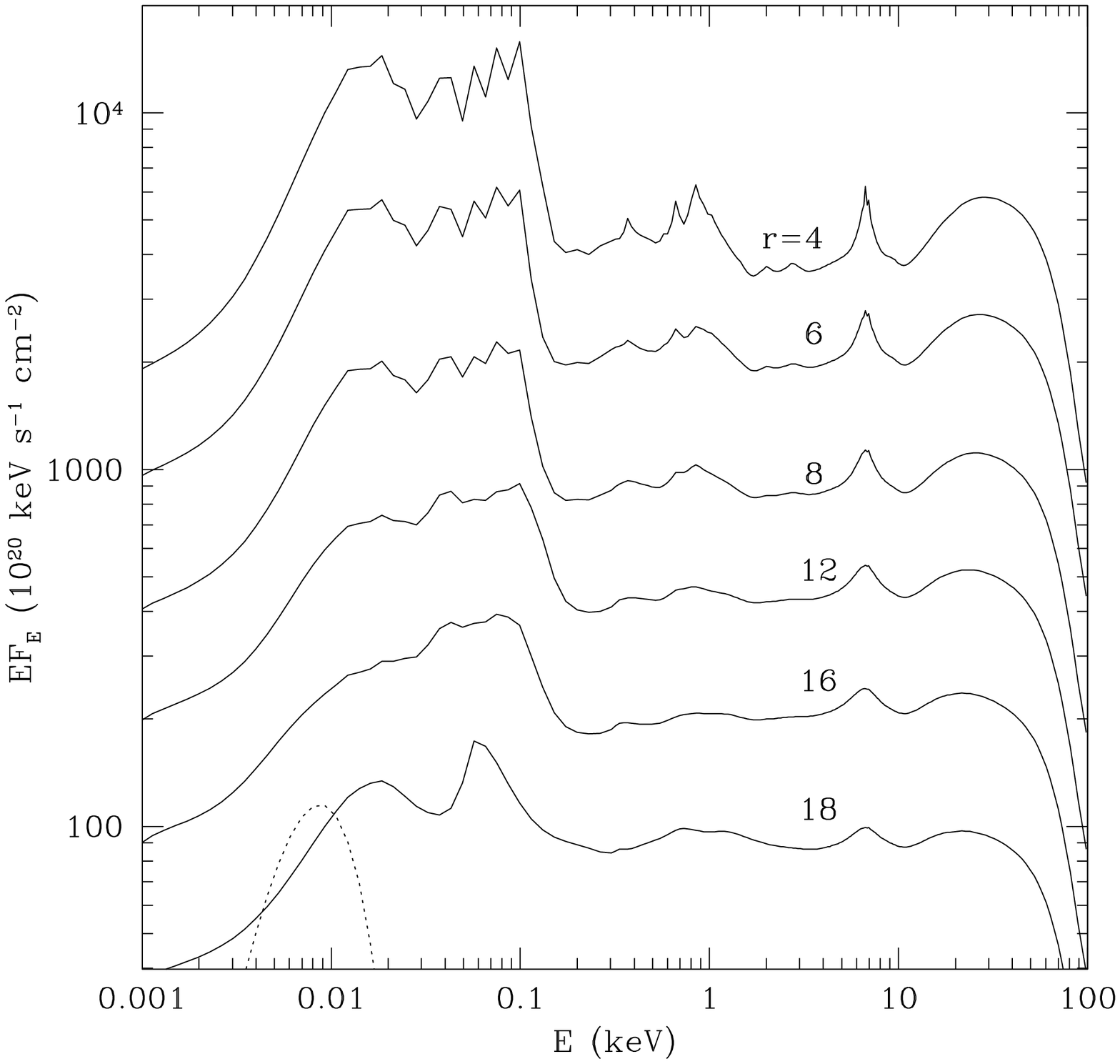,width=0.50\textwidth,silent=}}
\centerline{\psfig{figure=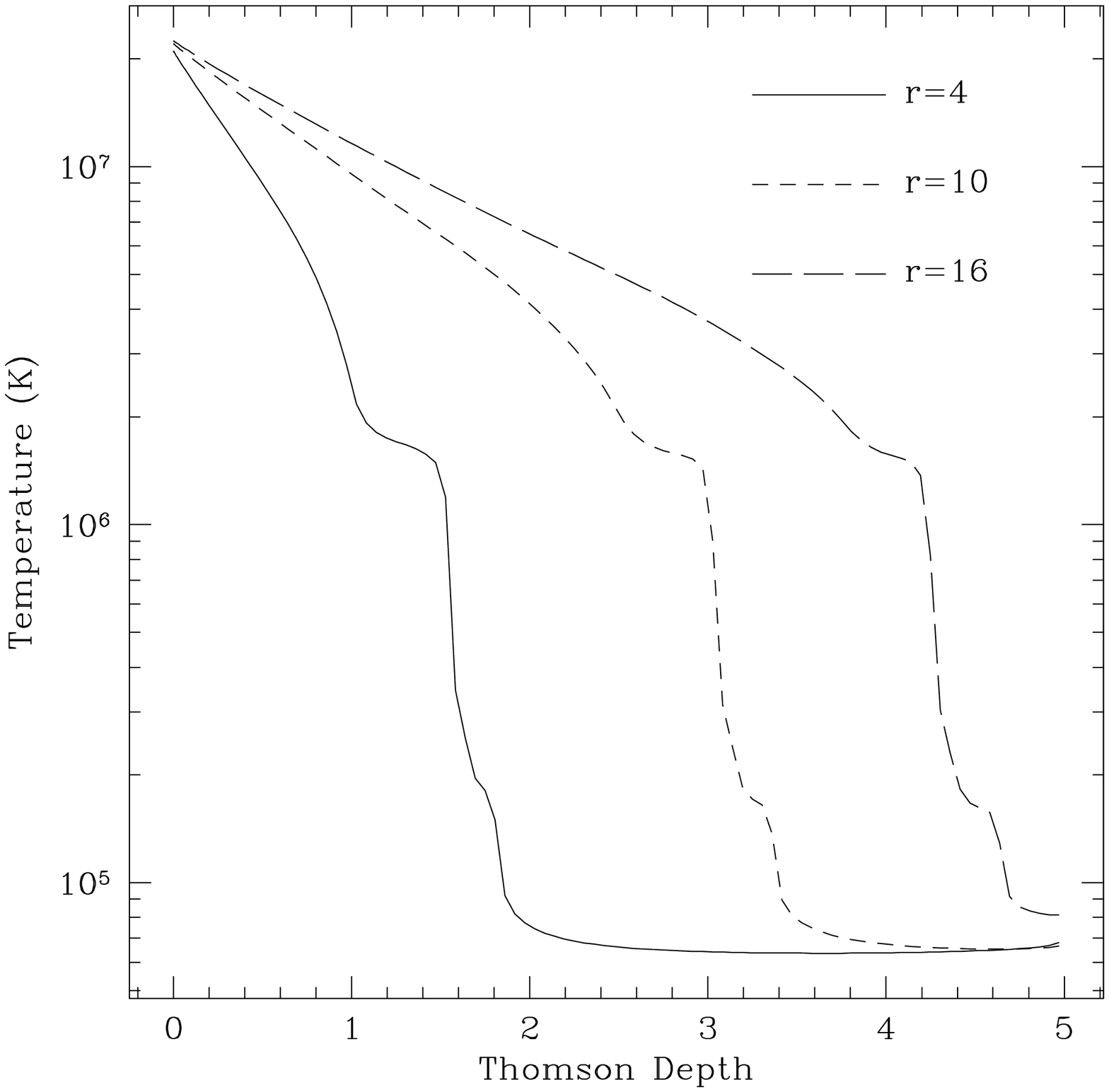,width=0.50\textwidth,silent=}}
\caption{{\it Top}: Reflection spectra for models where the radius at
which the illumination occurs has been varied. All the other
parameters have the same values as in the canonical model. The spectra
have been offset vertically for clarity. The dotted curve shows the
spectrum from the disc (when $r=4$) that is incident on the
base of the atmosphere. The model with $r=18$ ionizes through five Thomson depths, so a thicker atmosphere with $\tau_{\rm T}=10$ was used for this model. The spectra are more ionized as the radius increases, reflecting the decreasing the density of the disc (which is a result of the increasing height).  {\it
Bottom}: The temperature of the atmosphere as a function of Thomson
depth for three different values of $r$.}
\label{fig:radius}
\end{figure}

In Figure~\ref{fig:radius} we have plotted the reflection spectra and
the temperature structure for models at different radii along the
accretion disc, but with all other parameters held at their canonical
values (e.g., $F_x = 10^{15}$~\ergcms).
As the radius
increases, the height of the disc increases (eq.~\ref{eq:height}) which drops the density, and
the amount of soft flux emitted by the disc decreases
(eq.~\ref{eq:softflux}).  As a result of the decreasing density, the atmosphere is more ionized as the radius increases. In fact, when $r=18$ the radiation ionized through 5 Thomson depths, and so a $\tau_{\rm T}=10$ layer was used for this model.

\subsection{Varying the black hole mass}
\label{sub:mass}
\begin{figure}
\centerline{\psfig{figure=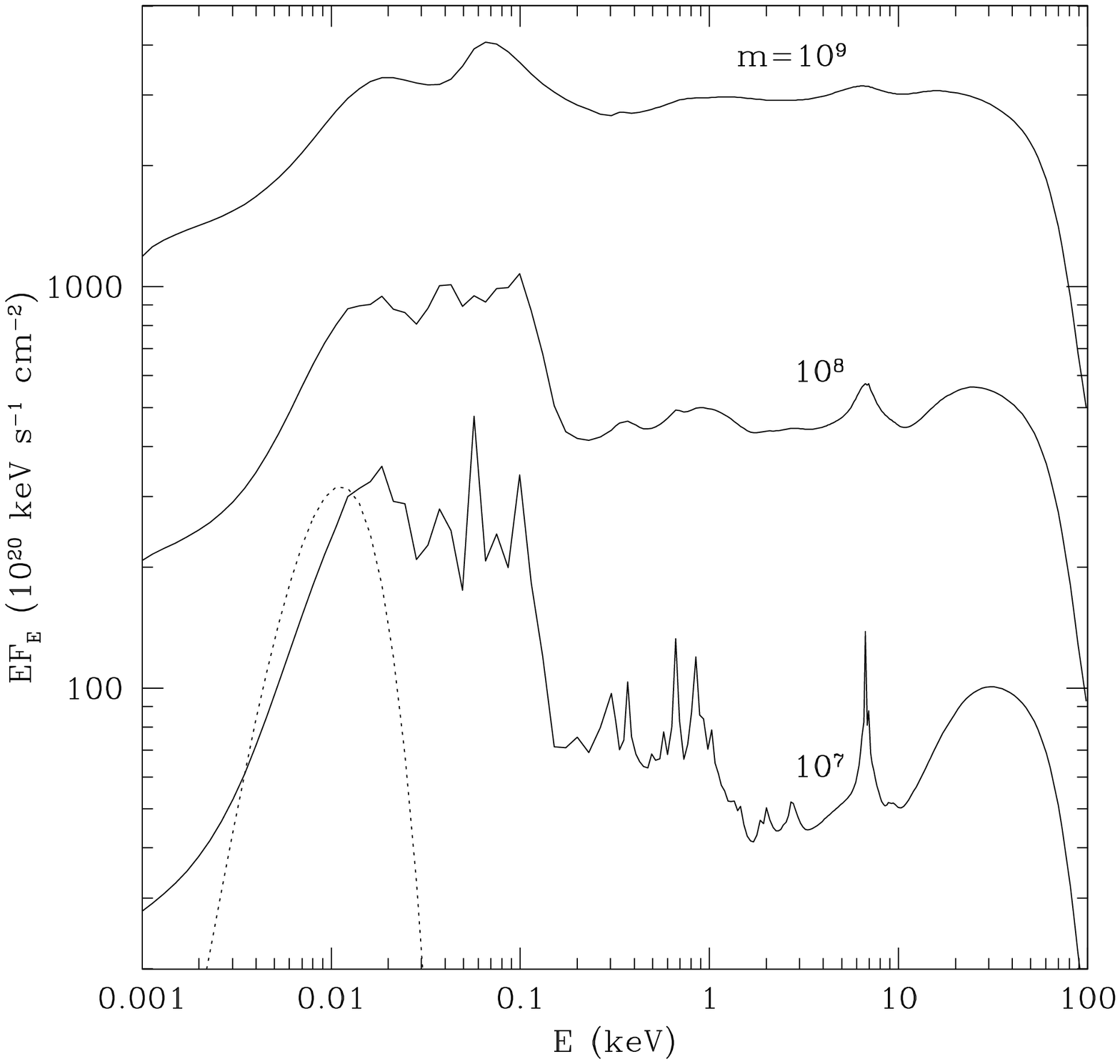,width=0.50\textwidth,silent=}}
\centerline{\psfig{figure=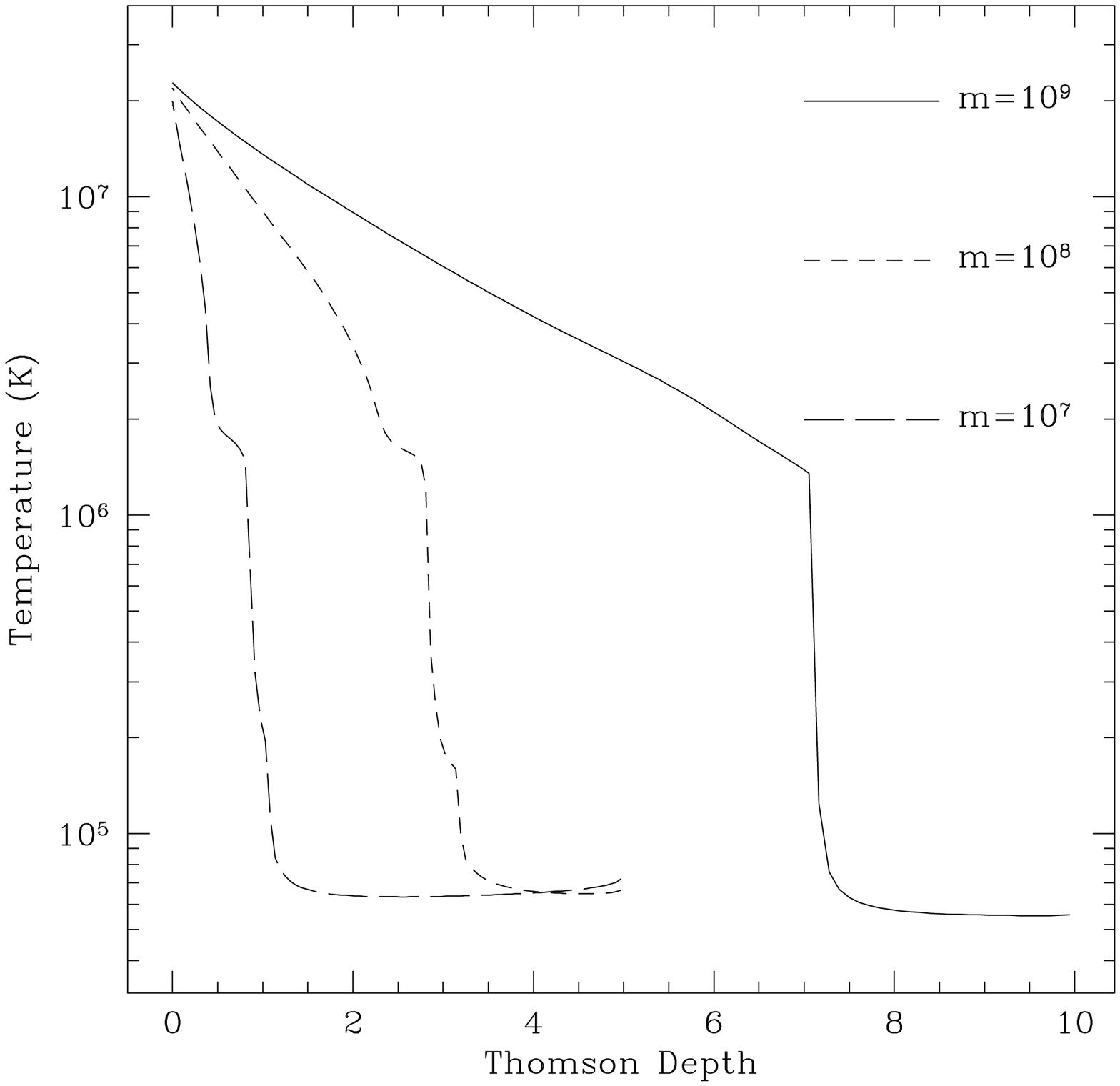,width=0.50\textwidth,silent=}}
\caption{{\it Top}: Reflection spectra for models where the black hole
mass has been changed. All the other parameters have the same values
as in the canonical model. The spectra have been offset vertically for
clarity. The dotted curve shows the spectrum from the disc
(when $m=10^7$) that is incident on the base of the atmosphere. When $m=10^9$, the density of the disc was so low that the incident radiation ionized through 5 Thomson depths, and so a deeper layer with $\tau_{\rm T}=10$ was used for this model. These reflection spectra show the differences of the changing density with black hole mass. {\it
Bottom}: The temperature of the atmosphere as a function of Thomson
depth for three different values of $m$.}
\label{fig:mass}
\end{figure}

In Figure~\ref{fig:mass} we plot
reflection spectra for the cases when the black hole mass has been
increased to 10$^9$~M$_{\odot}$ and decreased to
10$^7$~M$_{\odot}$. 
The model with the lowest black hole mass is not able to ionize as far
into the atmosphere as compared with the ones with larger masses. This
is because the disc becomes thinner as the black hole mass decreases
(see eq.~\ref{eq:height}), so the top layer of the atmosphere is
actually much denser in the low $m$ cases than in the high $m$
cases. Specifically, in the $m=10^9$ model, we find the density at the
top of the atmosphere to be $\sim 10^8$~cm$^{-3}$, while in
the $m=10^7$ model, this density has increased to $\sim 6 \times
10^9$~cm$^{-3}$. Therefore, the exact same ionizing spectrum cannot
penetrate as far into the layer. Furthermore, the larger density will
increase the cooling rates which will contribute to the temperature
transition occurring closer to the surface. As a result, the reflection
spectrum for the smaller black hole masses show stronger features from
neutral reflection (but still show ionized \fe\ lines). For the $m=10^9$ model an atmosphere with $\tau_{\rm T}=10$ had to be used to fully realize this situation.

As is discussed in Sect.~\ref{sect:limits}, our
code has difficulty if the thermal transition occurs at a Thomson
depth $\la 0.2$. As the black hole mass decreases, the disc becomes
denser, so the same amount of ionizing flux will not penetrate as
far. Therefore, we cannot model systems with black hole masses much
smaller than 10$^7$~M$_{\odot}$ without ramping up the illuminating
flux. In this section we are only interested in comparing the results
when one parameter is varied at a time, so using $F_x=10^{15}$~\ergcms\ 
restricts us to a limited range of black hole mass. 

\subsection{Varying the accretion rate}
\label{sub:mdot}
\begin{figure}
\centerline{\psfig{figure=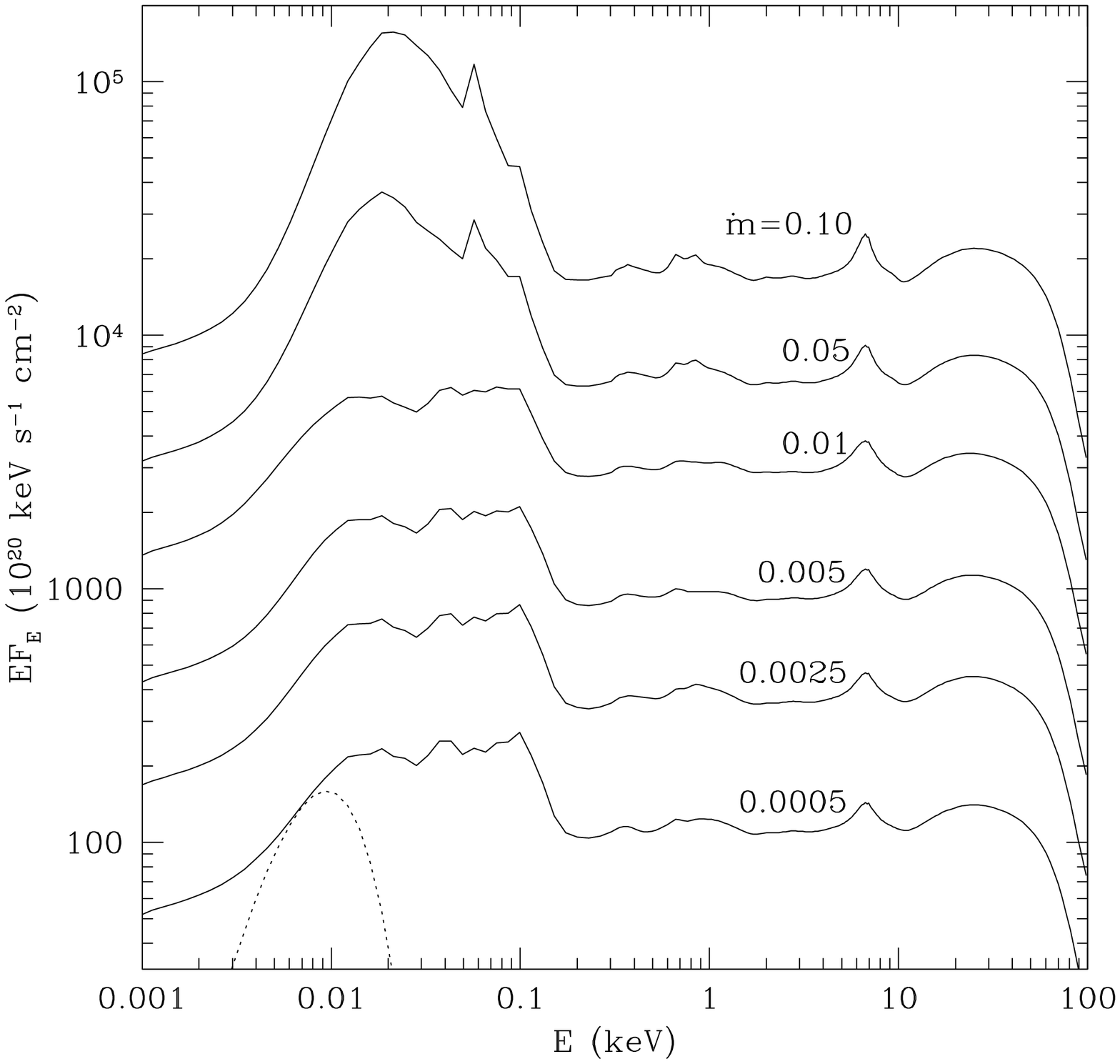,width=0.50\textwidth,silent=}}
\centerline{\psfig{figure=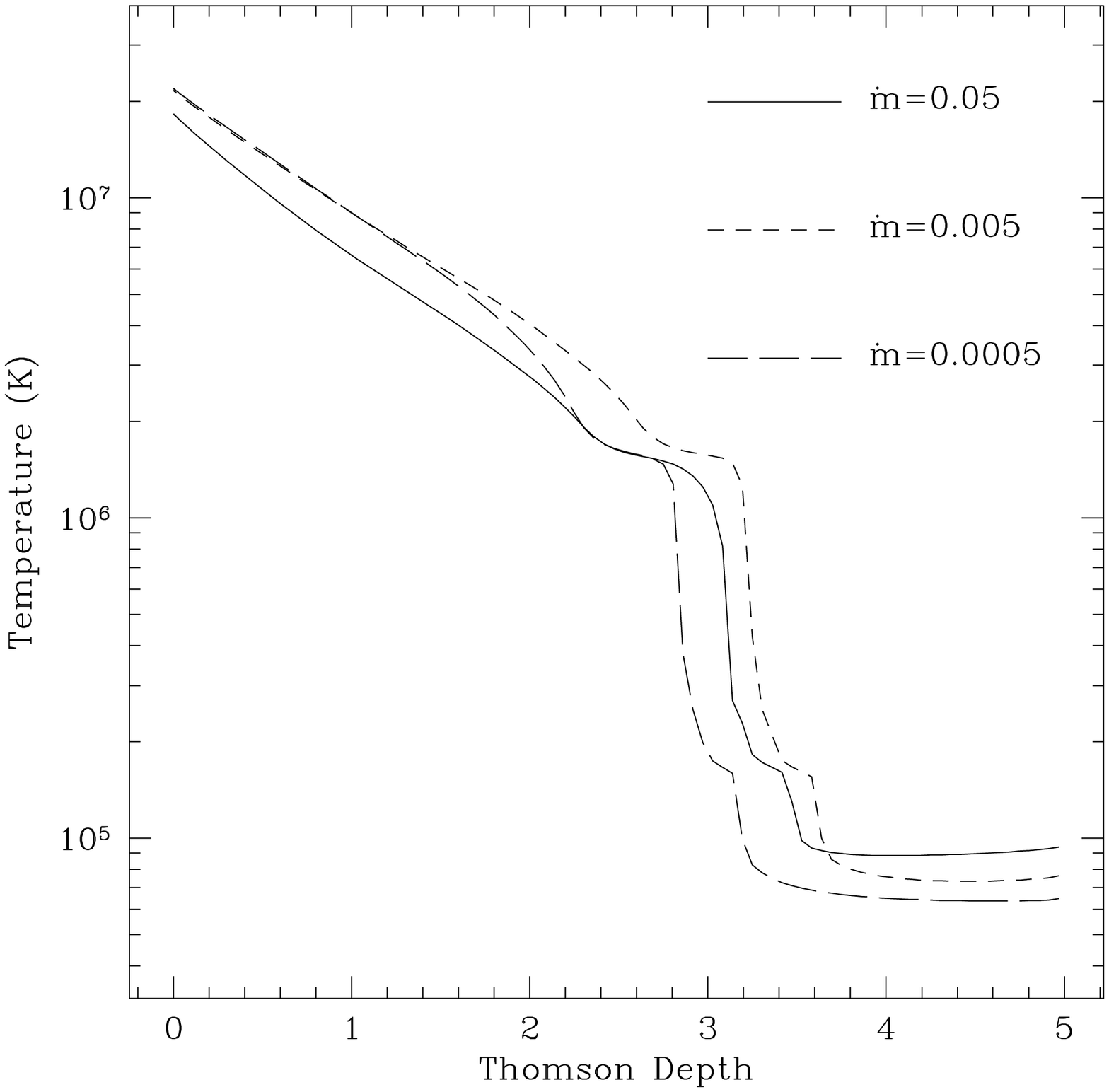,width=0.50\textwidth,silent=}}
\caption{{\it Top}: Reflection spectra for models where the accretion
rate has been varied. All the other parameters have the same values as
in the canonical model. The spectra have been offset vertically for
clarity. The dotted curve shows the spectrum from the disc (when $\dot
m=0.005$) that is incident on the base of the atmosphere. The spectra
for $\dot m=0.05$ and 0.10 were calculated with the
radiation pressure dominated boundary conditions from Merloni \etal\ (2000). {\it Bottom}: The
temperature of the atmosphere as a function of Thomson depth for three
different values of $\dot m$. The small temperature increase at high
Thomson depth in the $\dot m=0.05$ model is a result of the large soft
flux emanating from the disc.}
\label{fig:mdot}
\end{figure}

We have varied the accretion rate over three orders of magnitude, and
the results are displayed in Figure~\ref{fig:mdot}. As the accretion
rate is increased, the amount of soft disc radiation increases.
However, our boundary condition, the height $H$ of the base of the
atmosphere, is appropriate for only a gas-pressure dominated disc, and
so it does not rise as rapidly with $\dot m$ as is needed to
compensate for the increasing radiation pressure. Indeed, increasing
the accretion rate much beyond $\dot m=0.01$ results in a large
density inversion in the lower regions of the atmosphere, as the gas
has to push downwards to assist the weak gravity (due to low $z/r$)
and balance the strong radiation pressure from the thermal disc
emission. Therefore, for
$\dot m=0.05$ and 0.1 radiation pressure boundary conditions from Merloni \etal\ (2000) were used instead of the gas pressure dominated ones.

We find that there is not much difference between the reflection
spectra as the accretion rate varies. This result indicates that,
aside from the height of the disc, the internal structure of the
atmosphere is not greatly affected by changing the accretion
rate. However, the soft flux from the disc does increase with $\dot m$
and becomes comparable to the illuminating flux when $\dot m \approx
0.05$. At this point, a strong O~{\sc v}
recombination line at about 0.05~\kev\ is observed in the spectra. Also, the bremsstrahlung flux
between 0.01 and 0.1~\kev\ increases with $\dot m$, as the strong soft
disc radiation is thermalized by the ionized gas.  However, the hard
X-ray spectra above 1~\kev\ displays little variation as the accretion
rate is changed.
 
\section{Comparison with previous work}
\label{sect:compare}
Previous studies of accretion disc atmospheres that incorporated
hydrostatic balance have been carried out by Raymond (1993), Ko \&
Kallman (1994), R\'{o}\.{z}a\'{n}ska \& Czerny (1996), and
R\'{o}\.{z}a\'{n}ska (1999). The first two papers dealt with the ultraviolet
and soft X-ray emission from Low-Mass X-ray Binaries, while
the next two, although dealing with active galactic nuclei
(AGN), did not include detailed calculations of the radiative transfer
of spectral features. Therefore, it is not possible to compare our
results with these earlier papers.

Recently, Nayakshin \etal\ (2000) have calculated the X-ray reflection
spectrum from an ionized AGN accretion disc atmosphere.  Their
calculations included a comparable amount of physics as ours, so in
principle a comparison between the two approaches could be
made. However, the reflection spectra presented in their paper were
calculated by artificially varying the $A$ parameter while keeping all
the other parameters fixed. This was done to try to account for local
dynamical effects that might occur if the disc was illuminated by
bright magnetic flares (e.g., Galeev, Rosner \& Vaiana 1979; Haardt,
Maraschi \& Ghisellini 1994; Svensson 1996). In our calculations, $A$
is fixed once we have chosen all the parameters. As a result, there
are no common calculations between the two approaches and a comparison
cannot be made at this time.

\begin{figure*}
\begin{minipage}{150mm}
\centerline{\psfig{figure=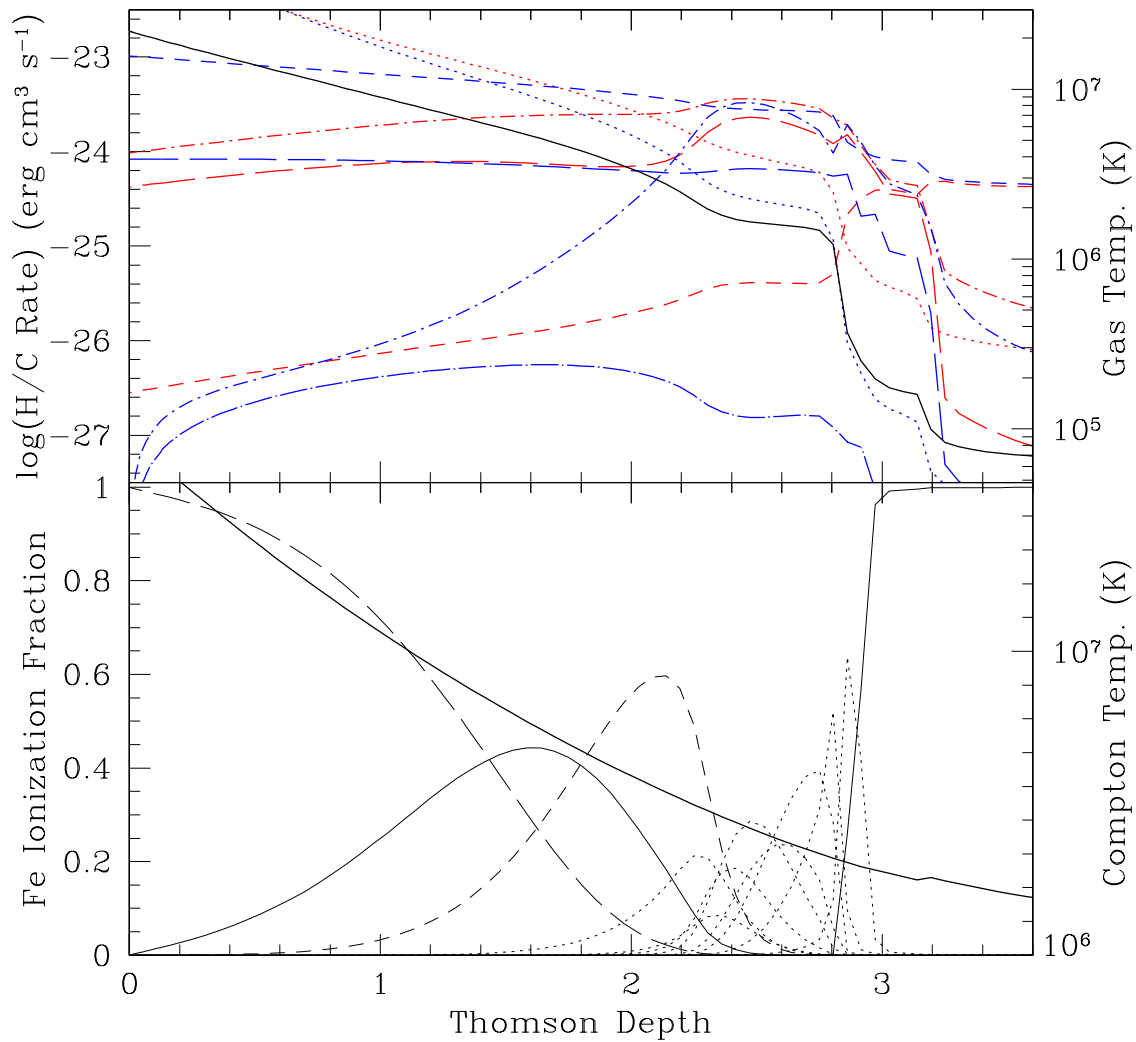,width=\textwidth,silent=}}
\caption{{\it Top}: (H)eating and (C)ooling rates for various
processes in the canonical model plotted versus the range of Thomson
depth that covers the transition region. The solid black curve denotes
the gas temperature. The rates are denoted as follows: red
curves=heating processes, blue curves=cooling processes,
dotted=Compton processes, short dash=bremsstrahlung, long
dash=recombination, dot-short dash=photoelectric heating/line cooling,
and dot-long dash=three-body. {\it Bottom}: The dark solid line plots
the Compton temperature versus Thomson depth. The other curves
illustrate the Fe ionization structure over this region. The long
dashed curve denotes completely ionized iron, the solid curve closer
to the surface shows hydrogenic iron, the short dashed line plots
helium-like iron, the dotted curves correspond to Fe~{\sc xvii--xxv},
and the solid curve near the bottom of the transition zone plots
``neutral'' Fe~{\sc xvi}.}
\label{fig:rates}
\end{minipage}
\end{figure*}

The results of these new calculations are more complicated to
interpret than earlier constant density ones where the reflection
spectrum depends only on the ionization parameter of the gas. Here,
the ionization parameter changes with depth as the density
changes. Therefore, different regions of the reflection spectrum will
arise from different depths within the layer. The temperature and
ionization structures of our canonical model, plus the strengths of
the processes which setup those structure, are shown in
Figure~\ref{fig:rates}. The top panel plots the heating and cooling
rates for various processes over the transition region for the
canonical model (see the caption for identification of the various
curves). The bottom panel plots both the Fe ionization structure and
the Compton temperature over the same region.
At the top of the atmosphere, Compton heating and cooling are the
dominant processes, Fe is completely ionized, and the gas temperature
is comparable to the Compton temperature. However, as we move into the
layer, the Compton temperature immediately starts dropping since
ionizing photons are being reflected by the ionized top and are not
making it further into the layer.  Hydrogenic iron begins to build up
and is the dominant species at a Thomson depth of about 1.4. Note that
the temperature plateau has not yet been reached, but the gas is
hotter then the Compton temperature. At this point Compton cooling is
surpassed by bremsstrahlung cooling because of the increasing
density. This cooling quickly exceeds the Compton heating rate, and
the gas temperature takes its first sharp turn downwards. Helium-like
Fe quickly builds up to be the dominant species (it is the line at
6.7~\kev\ from this species that is the most prominent line in the
reflection spectrum) just before $\tau_{\rm T}=2.2$. The plateau is
still not reached yet, but photoelectric heating from the quickly
recombining ions has become the dominant form of heating. Free-free
cooling is still dominating the cooling, but line cooling is quickly
gaining and dominates just past a Thomson depth of 2.3. This is when
the stable temperature plateau is found. It is kept stable by the
balance of photoelectric heating of the various ionization states of
Fe and its subsequent line cooling plus free-free cooling.  Finally,
the ionizing photon density drops to the point where everything can
recombine and we are left with neutral material with free-free heating
and cooling dominating.

\section{Fitting the new models with constant density ones}
\label{sect:const-dens}
Previous fits to X-ray spectra that employed reflection models were
limited to ones where the density was assumed to be constant within
the atmosphere. The most widely used publicly available models have
been the {\sc pexriv} (for ionized reflection) and {\sc pexrav} (for
neutral reflection) models of Magdziarz \& Zdziarski (1995) that are
included as part of the {\sc xspec} spectral fitting package. The
constant density models of Ross \& Fabian (1993) have also been used
to fit X-ray spectra (e.g., Ballantyne, Iwasawa \& Fabian 2001). Now
that variable density models are available, it is instructive to fit
them with the older, constant density ones. This exercise will
illustrate the differences between the variable density models and
constant density ones. It will also point out possible errors in
analysis that could have been made as a result of fitting constant
density models to spectral data. To isolate these effects, we have not
added any further spectral complexity (i.e., warm absorbers, soft
excess, or neutral absorption) to the model spectra.

We begin by generating simulated spectra using the results of the
calculations.  This is accomplished by first adding the reflection
spectrum to the illuminating spectrum (so that the reflection
fraction, $R$, is unity), and then using the `fakeit' command in {\sc
xspec}~v.11 to create simulated spectra with a certain exposure
time. To examine the results of fitting the models over different
energy ranges, we used the {\it RXTE} and the {\it ASCA} SIS0 response
matrices from our study of Ark~564 (Ballantyne \etal\ 2001) to produce
two different versions of each simulated spectrum. The normalization
of the models and the exposure times for each spectrum-response matrix
pair are shown in Table~\ref{table:simulated-params}, and were chosen
so that there were at least 20 counts per bin (which allows the use of
the $\chi^2$ statistic as a goodness of fit parameter). The
normalizations are so small because the model predicts flux in units
of emitting area. The values chosen here are typical of ones that are 
obtained when fitting real data. 
\begin{table*}
\begin{minipage}{100mm}
\caption{Normalizations and exposure times assumed for the simulated 
spectra that are analyzed in this section.}
\label{table:simulated-params}
\begin{tabular}{ccccc}
 &\multicolumn{2}{c}{{\it RXTE}} &\multicolumn{2}{c}{{\it ASCA}} \\ \cline{2-5}
$\Gamma / F_x^a$ & $\log$(Norm.$^b$) & Exp. Time$^c$ & $\log$(Norm.$^b$) & Exp. Time$^c$ \\ \hline
2.2 & $-27$ & 200 & $-25$ & 175 \\
2.1 & $-27$ & 175 & $-25$ & 150 \\
2.0 & $-27$ & 100 & $-25$ & 100 \\
1.9 & $-27$ & 100 & $-25$ & 100 \\
1.8 & $-27$ & 100 & $-25$ & 100 \\
1.7 & $-27$ & 100 & $-25$ & 100 \\
1.6 & $-27$ & 100 & $-25$ & 100 \\ \hline
$10^{16}$ & $-27$ & 100 & $-26$ & 100 \\
$5 \times 10^{15}$ & $-27$ & 100 & $-25$ & 100 \\
$5 \times 10^{14}$ & $-26$ & 100 & $-25$ & 400 \\
$10^{14}$ & $-26$ & 100 & $-24$ & 200 \\
\end{tabular}
\medskip

$^a$ units of \ergcms

$^b$ units of photons cm$^{-2}$ s$^{-1}$

$^c$ units of kiloseconds
\end{minipage}
\end{table*}
We then fit the spectra using the chosen model over three different
energy ranges: 3--20~\kev\ (using the {\it RXTE} response matrix),
0.5--10~\kev\ and 1--10~\kev\ (using the {\it ASCA} SIS0 response
matrix). It is worth pointing out that because of the randomness in
generating these simulated spectra (since `fakeit' includes random
photon noise in each spectrum), the actual numerical values of the
fits are not solid results. However, trends in the fits and whether or
not a spectrum is fit `well' or fit `poorly' are valid results, and
can be used to draw conclusions.

We first fit the {\sc pexriv} and {\sc pexrav} models to our variable
density models which have had the photon index varied from 2.2 to 1.6
(the other model parameters are the same as the canonical model; see
Table~\ref{table:models}). As seen in Fig.~\ref{fig:gamma}, these
spectra cover the range from highly ionized and featureless
($\Gamma=1.6$), to ones which have more pronounced spectral features
($\Gamma=2.2$). Since {\sc pexriv} and {\sc pexrav} do not include any
line emission, each fit included a Gaussian to model the \fe\
line. The central energy of the Gaussian was fixed at 6.7~\kev\, which
is the energy of the \fe\ line in our model spectra, the width
was fixed at 0.5~\kev, and the normalization was allowed to vary. It
is worth pointing out that in these models there is no physical
connection between the strength and energy of the \fe\ line and the
edge that appears in the continuum. This is one of the chief
weaknesses of using the {\sc pexriv+gauss} or {\sc pexrav+gauss}
models to fit data which includes an \fe\ line. The continuum
parameters were set as follows: cutoff energy of power-law=400~\kev,
inclination angle=30$^{\circ}$, abundances=solar, and ({\sc pexriv}
only) temperature of disc=10$^6$~K. The results of these fits are
shown in Table~\ref{table:pexrivfits}.
\begin{table*}
\begin{minipage}{180mm}
\caption{Results of fitting the {\sc pexriv} and {\sc pexrav} models
to our new models (incident+reflected). These models were computed
with $m=10^8$, $\dot m$=0.001, $F_x = 10^{15}$~\ergcms, $r=9$,
$i=54\fdg7$, and had the photon index varied from 2.2 to 1.6. A
Gaussian line was included in all fits to model the \fe\
complex. The energy of the line was fixed at 6.7~\kev\ and the
intrinsic width was fixed at 0.5~\kev; however, the normalization was
free to vary. Further details on the parameters are discussed in the text.}
\label{table:pexrivfits}
\begin{tabular}{ccccccccccccccccccc}
 &\multicolumn{6}{c}{{\it RXTE} (3--20~\kev)} &\multicolumn{6}{c}{{\it
 ASCA} (0.5--10~\kev)} &\multicolumn{6}{c}{{\it ASCA} (1.0--10~\kev)}
 \\ \cline{2-19} &\multicolumn{3}{c}{{\sc pexriv}- 40 dof}
 &\multicolumn{3}{c}{{\sc pexrav}- 41 dof} &\multicolumn{3}{c}{{\sc
 pexriv}- 642 dof}&\multicolumn{3}{c}{{\sc pexrav}- 643 dof}
 &\multicolumn{3}{c}{{\sc pexriv}- 608 dof} &\multicolumn{3}{c}{{\sc
 pexrav}- 609 dof} \\
$\Gamma_{model}$ & $R$ & $\Gamma$ & $\chi^2$ & $R$ & $\Gamma$ &
 $\chi^2$ & $R$ & $\Gamma$ & $\chi^2$ & $R$ & $\Gamma$ &
 $\chi^2$ & $R$ & $\Gamma$ & $\chi^2$ & $R$ & $\Gamma$ & $\chi^2$ \\ \hline
2.2 & 0.04 & 2.21 & 50 & 0.0 & 2.20 & 50 & 1.63 & 2.39 & 770 & 1.75 & 2.39 & 771 & 1.38 & 2.41 & 628 & 3.27 & 2.43 & 628 \\
2.1 & 0.0 & 2.21 & 44 & 0.0 & 2.20 & 44 & 0.52 & 2.22 & 744 & 0.70 & 2.22 & 746 & 0.96 & 2.24 & 673 & 1.1 & 2.23 & 677 \\
2.0 & 0.0 & 2.00 & 42 & 0.0 & 1.99 & 42 & 0.88 & 2.13 & 671 & 1.3 & 2.13 & 672 & 0.68 & 2.12 & 588 & 2.1 & 2.16 & 611 \\ 
1.9 & 0.0 & 1.93 & 40 & 0.0 & 1.93 & 40 & 0.24 & 1.99 & 681 & 0.31 & 1.99 & 683 & 0.17 & 1.98 & 615 & 0.47 & 2.00 & 621 \\
1.8 & 0.17 & 1.82 & 55 & 0.16 & 1.83 & 56 & 0.39 & 1.90 & 644 & 0.69 & 1.91 & 652 & 0.56 & 1.92 & 595 & 0.70 & 1.90 & 605 \\
1.7 & 0.06 & 1.70 & 35 & 0.01 & 1.68 & 35 & 0.0 & 1.80 & 826 & 0.0 & 1.80 & 826 & 0.11 & 1.81 & 788 & 0.0 & 1.80 & 788 \\
1.6 & 0.0 & 1.61 & 41 & 0.0 & 1.60 & 41 & 0.24 & 1.71 & 701 & 0.06 & 1.70 & 705 & 0.0 & 1.68 & 651 & 0.0 & 1.68 & 651 \\
\end{tabular}
\end{minipage}
\end{table*}
The first noticeable result from the fits is that, in almost all cases, the
{\sc pexriv} and {\sc pexrav} models do a good job fitting the spectra
in the 3--20~\kev\ range: the $\chi^2$ of the fit is low and the
derived value of the photon index is close to the true value.  This is
due to the fact that, aside from the \fe\ line and its associated
edge, the spectra are relatively smooth and featureless in this
region. Of course, the model can fit both the line and the edge, so
that the $\chi^2$ is quite low. Even though the {\sc pexrav} model is
appropriate only for neutral reflection, it still provides a good fit
to the simulated data.  This might be because the resolution of {\it
RXTE} is poor enough that the difference between a neutral and an
ionized Fe edge cannot be distinguished in noisy data, or the \fe\
normalization was adjusted to help in the fitting. Also, the fits tend
to give much lower reflection fractions than the true value of $R$
(which is unity). Furthermore, the 1-$\sigma$ errors on the derived
reflection fraction are quite large, which suggests that the lack of 
features in this
spectral region, combined with the poor resolution of \textit{RXTE},
makes it difficult to constrain the amount of reflection in a spectrum
(this is also true for the Ross \& Fabian (1993) models; see
Table~\ref{table:rvarfits}). 

Moving to the {\it ASCA} energy ranges, we find that both
the {\sc pexriv} and the {\sc pexrav} models have greater difficulty providing adequate
fits to the data in the 0.5--10~\kev\ energy range: often the value of $\Gamma$ is
highly overestimated. This behaviour also persists in the 1--10~\kev\
region. Fig.~\ref{fig:gamma} shows that below about 2~\kev\
the reflection spectrum begins to steepen. The {\sc pexriv} and {\sc
pexrav} models do not have enough spectral curvature to account for
this change and therefore must
steepen the power-law in order to fit these low energy data. As a
result the model will underestimate the spectrum at the high energy
end, and must increase the amount of reflection to increase the quality
of the fit. The ionization state of the reflection spectrum increases as
$\Gamma$ decreases and therefore the amount of reflection the model
requires in order to obtain a best-fit will decrease. Ignoring the
data below 1~\kev\ will remove some of the
steepening, and the results obtained from {\sc pexriv} and {\sc
pexrav} improve slightly. The above results show that {\sc pexriv} and {\sc
pexrav} cannot adequately describe a reflection spectrum that arises
from a highly ionized disc. 

We then proceeded to fit the simulated spectra with the constant
density models of Ross \& Fabian (1993). These models, of course,
contain the same atomic and ionization physics as the ones presented
in this paper. Along with the models with the different values of
$\Gamma$, we also fit simulated spectra for the models with differing
amounts of illuminating flux (and $\Gamma=1.9$). The spectra were fit
two different times: first, with the reflection fraction of the model
fixed at unity (to investigate the dependence on the ionization
parameter $\xi=4 \pi F_x/n_{\rm H}$), and, second, when the
reflection fraction was allowed to be fit.  The results of the fits
are presented in Tables~\ref{table:r1fits} and~\ref{table:rvarfits},
respectively.
\begin{table*}
\begin{minipage}{115mm}
\caption{Results of fitting the constant models of Ross \& Fabian
(1993) to our new models (incident+reflected). The models in the top
part of the table were computed with $m=10^8$, $\dot m$=0.001, $F_x =
10^{15}$~\ergcms, $r=9$, $i=54\fdg7$, and had the photon index
varied from 2.2 to 1.6. The models in the bottom part of the table
were computed with $\Gamma=1.9$, and had the illuminating flux varied
(all other parameters were unchanged). The reflection fraction in the
constant density models was fixed at unity. For more details see the
text.}
\label{table:r1fits}
\begin{tabular}{cccccccccc}
 &\multicolumn{3}{c}{{\it RXTE} (3--20~\kev)} &\multicolumn{3}{c}{{\it
 ASCA} (0.5--10~\kev)} &\multicolumn{3}{c}{{\it ASCA} (1.0--10~\kev)}
 \\ \cline{2-10}
$\Gamma_{model} / F_x^a$ & $\log \xi$ & $\Gamma$ & $\chi^2/42$ & $\log \xi$
 & $\Gamma$ & $\chi^2/644$ & $\log \xi$ & $\Gamma$ & $\chi^2/610$ \\ \hline
2.2 & 3.80 & 2.11 & 51 & 4.50 & 2.26 & 804 & 3.83 & 2.23 & 629 \\
2.1 & 4.17 & 2.11 & 47 & 4.70 & 2.12 & 757 & 4.28 & 2.13 & 673 \\
2.0 & 3.91 & 1.91 & 46 & 6.0 & 1.99 & 671 & 3.94 & 2.02 & 598 \\
1.9 & 3.99 & 1.86 & 44 & 4.90 & 1.85 & 678 & 4.29 & 1.89 & 613 \\
1.8 & 4.10 & 1.72 & 54 & 4.51 & 1.71 & 650 & 4.05 & 1.75 & 606 \\
1.7 & 3.73 & 1.59 & 37 & 4.59 & 1.58 & 851 & 4.72 & 1.58 & 815 \\
1.6 & 4.11 & 1.51 & 44 & 4.71 & 1.50 & 898 & 5.02 & 1.50 & 717 \\ \hline
$10^{16}$ & 4.47 & 1.86 & 40 & 5.12 & 1.81 & 660 & 4.20 & 1.86 & 628 \\
$5 \times 10^{15}$ & 4.21 & 1.84 & 42 & 4.83 & 1.86 & 812 & 4.05 & 1.89 & 703 \\
$5 \times 10^{14}$ & 3.92 & 1.80 & 38 & 4.97 & 1.87 & 770 & 4.05 & 1.91 & 675 \\
$10^{14}$ & 3.72 & 1.77 & 34 & 4.36 & 1.92 & 887 & 3.67 & 1.93 & 729 \\
\end{tabular}
\medskip

$^a$ units of \ergcms
\end{minipage}
\end{table*}
Turning our attention to the fits where the reflection fraction, $R$,
was fixed at its true value of one, we find some interesting
results. First, as with the {\sc pexriv} and {\sc pexrav} model, the
best fits in a statistical sense were found in the 3--20~\kev\
\textit{RXTE} energy range. As before, in this region there are
only a few spectral features, so it easier for models to fit the
data. Note that this is quite a general result. To differentiate
between different models, or be confident about the model parameters,
it is preferable to fit data over an energy range where there will be
spectral features to fit against. In the three different energy ranges
considered in this experiment, we often find quite different values
for $\xi$. This is an important point to keep in mind when doing
spectral fitting. However, while the fits may be acceptable in a
$\chi^2$ sense in the \textit{RXTE} energy range, the derived values
of the photon index are consistently low. The fits are quite poor in the
0.5--10~\kev\ energy range, but do improve in the 1--10~\kev\
band. The computed values of $\Gamma$ are closer to the true values,
but still show erratic behaviour. 

These results drive home the point that the new hydrostatic models are
comprised of emission from regions with different ionization
parameters.  The ionization parameters that are found are, broadly
speaking, fairly high and quite variable. Therefore, trying to fit the spectra with a
model that has only one ionization parameter will generally lead to an
average value of $\xi$ or one which is from the dominant component in
the spectrum.  In these models, the most significant spectral feature
is from the \fe\ line of helium-like Fe {\sc xxv} at 6.7~\kev\ but it is weakened and broadened due to Compton scattering of the line photons. This
line is prominent in constant density models at values of $\xi$
above a few thousand (Ross \etal\
1999). Since this line is also prominent in the simulated spectra,
values of $\xi$ greater than a few thousand are typically
found by the fits. This illustrates the difficulty in fitting a single
ionization parameter model to a situation which has a mixture of
ionization states. 
\begin{table*}
\begin{minipage}{135mm}
\caption{Same as Table~\ref{table:r1fits}, but allowing the reflection
fraction to be fit.}
\label{table:rvarfits}
\begin{tabular}{ccccccccccccc}
 &\multicolumn{4}{c}{{\it RXTE} (3--20~\kev)} &\multicolumn{4}{c}{{\it
 ASCA} (0.5--10~\kev)} &\multicolumn{4}{c}{{\it ASCA} (1.0--10~\kev)}
 \\ \cline{2-13}
$\Gamma_{model} / F_x^a$ & $\log \xi$ & $\Gamma$ & $R$ & $\chi^2/41$ & $\log \xi$
 & $\Gamma$ & $R$ & $\chi^2/643$ & $\log \xi$ & $\Gamma$ & $R$ & $\chi^2/609$ \\ \hline
2.2 & 3.50 & 2.17 & 0.53 & 50 & 3.05 & 2.23 & 0.26 & 669 & 3.23 & 2.24 & 0.35 & 596 \\
2.1 & 3.71 & 2.18 & 0.29 & 45 & 3.02 & 2.13 & 0.15 & 695 & 3.05 & 2.14 & 0.17 & 655 \\
2.0 & 3.60 & 1.97 & 0.32 & 45 & 4.96 & 1.90 & 9.9 & 653 & 4.14 & 1.95 & 9.9 & 587 \\
1.9 & 3.21 & 1.94 & 0.14 & 41 & 4.92 & 1.85 & 0.93 & 678 & 4.26 & 1.90 & 0.78 & 612 \\
1.8 & 4.19 & 1.63 & 9.8 & 54 & 4.54 & 1.78 & 0.49 & 646 & 4.00 & 1.81 & 0.46 & 599 \\
1.7 & 3.09 & 1.69 & 0.15 & 35 & 5.03 & 1.75 & 0.13 & 828 & 3.40 & 1.78 & 0.05 & 791 \\
1.6 & 3.80 & 1.62 & 0.03 & 43 & 2.65 & 1.67 & 0.03 & 700 & 2.70 & 1.67 & 0.02 & 655 \\ \hline
$10^{16}$ & 4.47 & 1.86 & 1.06 & 40 & 6.0 & 1.87 & 0.33 & 653 & 4.10 & 1.93 & 0.22 & 615 \\
$5 \times 10^{15}$ & 4.25 & 1.83 & 1.33 & 42 & 4.56 & 1.75 & 4.5 & 766 & 4.15 & 1.86 & 1.7 & 696 \\    
$5 \times 10^{14}$ & 3.58 & 1.86 & 0.26 & 36 & 4.52 & 1.73 & 9.9 & 734 & 2.65 & 1.95 & 0.20 & 637 \\
$10^{14}$ & 3.03 & 1.86 & 0.21 & 29 & 4.40 & 1.80 & 9.9 & 841 & 2.65 & 1.96 & 0.31 & 619 \\
\end{tabular}
\medskip

$^a$ units of \ergcms
\end{minipage}
\end{table*}

When the reflection fraction is allowed to vary, the quality of the
fits increases significantly. In the \textit{ASCA} energy band, a
similar relationship between $R$ and $\Gamma$ to the one seen using
the {\sc pexriv} and {\sc pexrav} model is observed. Finally, this exercise emphasizes that the new hydrostatic 
reflection models can be qualitatively described as diluted versions of
the constant density ones.

\subsection{The $\bmath{R}$--$\bmath{\Gamma}$ correlation}
\label{sub:rgamma}
Zdziarski, Lubi\'{n}ski \& Smith (1999) have fit \textit{Ginga} spectra of
Seyfert galaxies and X-ray binaries with the {\sc pexrav} model, and
found that the amount of reflection correlated with the photon
index. While this was initially interpreted as a possible physical
correlation due to disc-corona feedback effects, it now seems
likely that much of this correlation is a result of systematic errors
in the standard data fitting procedures (e.g., Vaughan \& Edelson
2001; Chiang \etal\ 2001; Nandra \etal\ 2000). However, recent work by
Done \& Nayakshin (2001) suggested that ionization effects might also
contribute to the observed correlation. This may be expected as it is
difficult to distinguish in data the difference between a
highly-ionized disc which is strongly reflective and a disc which is
more neutral but weakly reflective. Done \& Nayakshin (2001) employed
a \textit{Ginga} response matrix, and found a $R$--$\Gamma$ correlation
when fitting the models of Nayakshin \etal\ (2000) with {\sc pexriv}.

Although there is some scatter we find similar trends in the \textit{ASCA} bands when using all three constant density models. The reason
for this was explained in the previous section and of course was due
to ionization effects. Therefore, we can conclude that perhaps part of
the $R$--$\Gamma$ correlation is due to the effects of ionized features
in the reflection spectrum.

\section{Limitations of current models}
\label{sect:limits}

The greatest uncertainty in the calculations presented in this paper
is the choice for the lower boundary condition of the layer.  We have
chosen to match the bottom height of the atmosphere to the expected
thickness of a gas-dominated disc. Alternatively, Nayakshin \etal\
(2000) matched the total pressure at the bottom of the atmosphere, but
had to assume the pressure followed a Gaussian drop-off from the disc
midplane. Both choices are equally valid (or equally invalid), as the
true physical structure and extent of accretion discs are not well
constrained. The choice of a gas-dominated boundary condition limits the
 applicability of the models to systems with $\dot m \approx 0.001$ at
 small radii, or with higher accretion rates at larger radii. These
 limits can be increased if we let a large amount of accretion power
 be dissipated in the corona (Svensson \& Zdziarski 1994). 

The significant computational limitation with our current code is that
the illuminating radiation must have a certain amount of ionizing
power.  In the context of the $A$ parameter of Nayakshin
\etal\ (2000), this limitation translates into a maximum $A$ value
($\approx 1$). The difficulties are due to the thermal transition
occurring too close to the surface of the atmosphere (say, $\tau_{\rm
T} \la 0.2$). When this situation arises, the majority of the gas in
the atmosphere is either fully or partially recombined, and, as the
model relaxes, a weak ionization front propagates inwards. It is
difficult for the code to get radiation out past the front since the
optical depth is so high (it is much easier when the gas is mostly
ionized and the gas recombines from the inside outwards). Moreover, in
these situations, the transition zone occurs in an optically-thin
region, and our diffusion treatment for radiative transfer becomes
inappropriate. In these cases, it is likely that the reflection
spectrum will be dominated by mainly neutral reflection features (much
like the ones produced by constant density models with low ionization
parameters).
 
\section{Discussion}
\label{sect:discuss}
The reflection spectra presented in Sect.~\ref{sect:res} show that,
for highly illuminated discs, ionized features are quite common over a
wide range of parameters. However, analyses of {\it ASCA} data of
samples of Seyfert~1 galaxies (Nandra \etal\ 1997a) and quasars (George
\etal\ 2000) show that ionized \fe\ lines are unusual, if not rare
(but see Reeves \& Turner 2000).  Typical values of $\Gamma$ for these
objects are around 1.9 or 2.0, and, by Fig.~\ref{fig:gamma}, we see
that ionized Fe lines are predicted by the models for these values of
the photon index. There are a number of possible explanations for this
discrepancy. One important consideration is that because of low
signal-to-noise data (especially for distant quasars), there is
usually a significant uncertainty in measuring the energy of the \fe\
line. Often it is difficult to rule out an ionized line. Second, if
the line emission is originating from material within a dozen or so
Schwarzschild radii from the central black hole, then gravitational
redshifting will be important, and an ionized line could be shifted so
that it would appear neutral. There are also possible physical
explanations for this result. Perhaps most discs are weakly
illuminated, denser, or they are being irradiated at large incidence
angles. In these cases, it is more likely that purely neutral reflection
will occur. We will have to await high signal-to-noise spectral data
in the region around the \fe\ line before these possible explanations 
can be tested.

Another interesting result from the calculations is that highly
ionized, featureless reflection spectra arise only when the photon
index of the illuminating power-law is less than 1.7. If simple
constant density reflection models are fit to such a spectrum (when it
has been added to the power-law component), they will lead to an
underestimate in the strength of the reflection component in the
sum. This is a possible explanation for the low reflection fractions
that have been observed in Galactic Black Hole Candidates (GBHCs),
when they are in their low/hard states (Gierli\'{n}ski \etal\ 1997;
Done \& \.{Z}ycki 1999). In this case, there would be no need to
appeal to a two-phase accretion flow to explain the observations. 

Nayakshin \& Kallman (2001) and Nayakshin (2000) presented evidence
that X-ray reflection results from reprocessing of radiation from
magnetic flares in the corona. The basis for this conclusion was that
the models of Nayakshin \etal\ (2000) predict that the equivalent
width of the {\it neutral} \fe\ line decreases as the illuminating
flux increases (as long as $F_{x} \gg F_{\rm disc}$, a requirement for
the magnetic coronal models). They compare this result to the observed
trend of decreasing \fe\ equivalent width with X-ray luminosity (the
``X-ray Baldwin effect''; Iwasawa \& Taniguchi 1993; Nandra \etal\
1997b; Reeves \& Turner 2000), and conclude that this is strong
evidence for the magnetic flare model as compared to the lamppost
model, which did not show this trend (since $F_{x} \ll F_{\rm
disc}$). Again, these coronal models had the relative strength of
gravity on the atmosphere artificially increased by a certain
factor. From Fig.~\ref{fig:fx}, we see that, as we increase $F_x$ over
two orders of magnitude (the same range as Nayakshin 2000), the
equivalent width of the \fe\ line drops slightly. However, it does not
go to zero, and it is unclear whether the line would disappear if the
illuminating flux continued to increase.  We do not disagree that
magnetic flares are a good model for the origin of the hard X-ray
emission in AGN, but the X-ray Baldwin effect is not a good test for
such a model. The \fe\ equivalent width is observed to drop once the
2--10~\kev\ luminosity surpasses $\sim 10^{45}$~erg~s$^{-1}$ (e.g.,
Nandra \etal\ 1997b). If these high luminosity objects have similar
black hole masses as their lower luminosity cousins, then their higher
luminosity might be caused by a higher accretion rate (perhaps above
$\dot M_{\rm Edd}$). At such high rates, the accretion disc will be
puffed up significantly due to the large radiation pressure from
within. In these cases, it is likely there will not be an
optically-thick surface for X-ray reprocessing, and the \fe\ line will
not appear in an X-ray spectrum.

\begin{figure}
\centerline{\psfig{figure=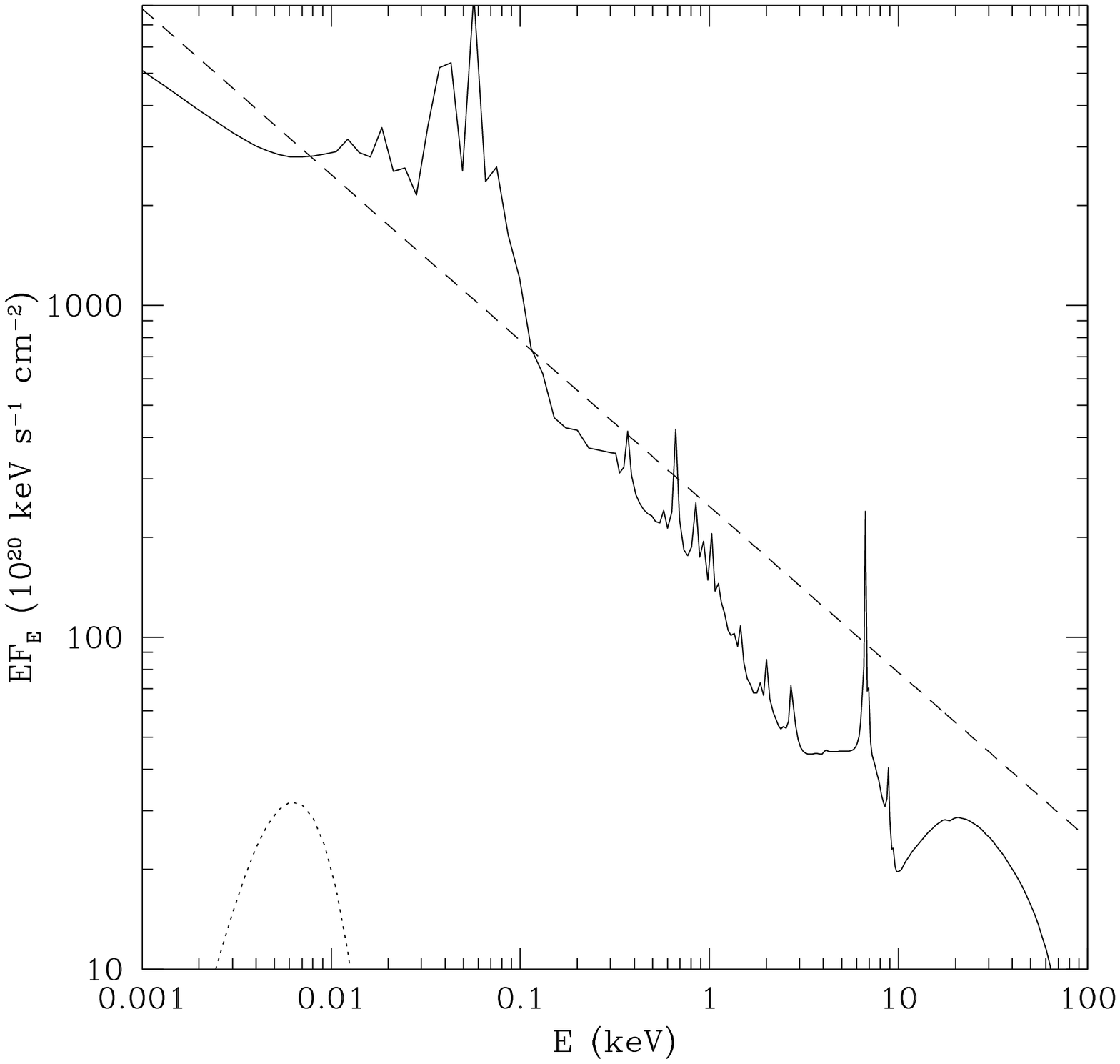,width=0.50\textwidth,silent=}}
\centerline{\psfig{figure=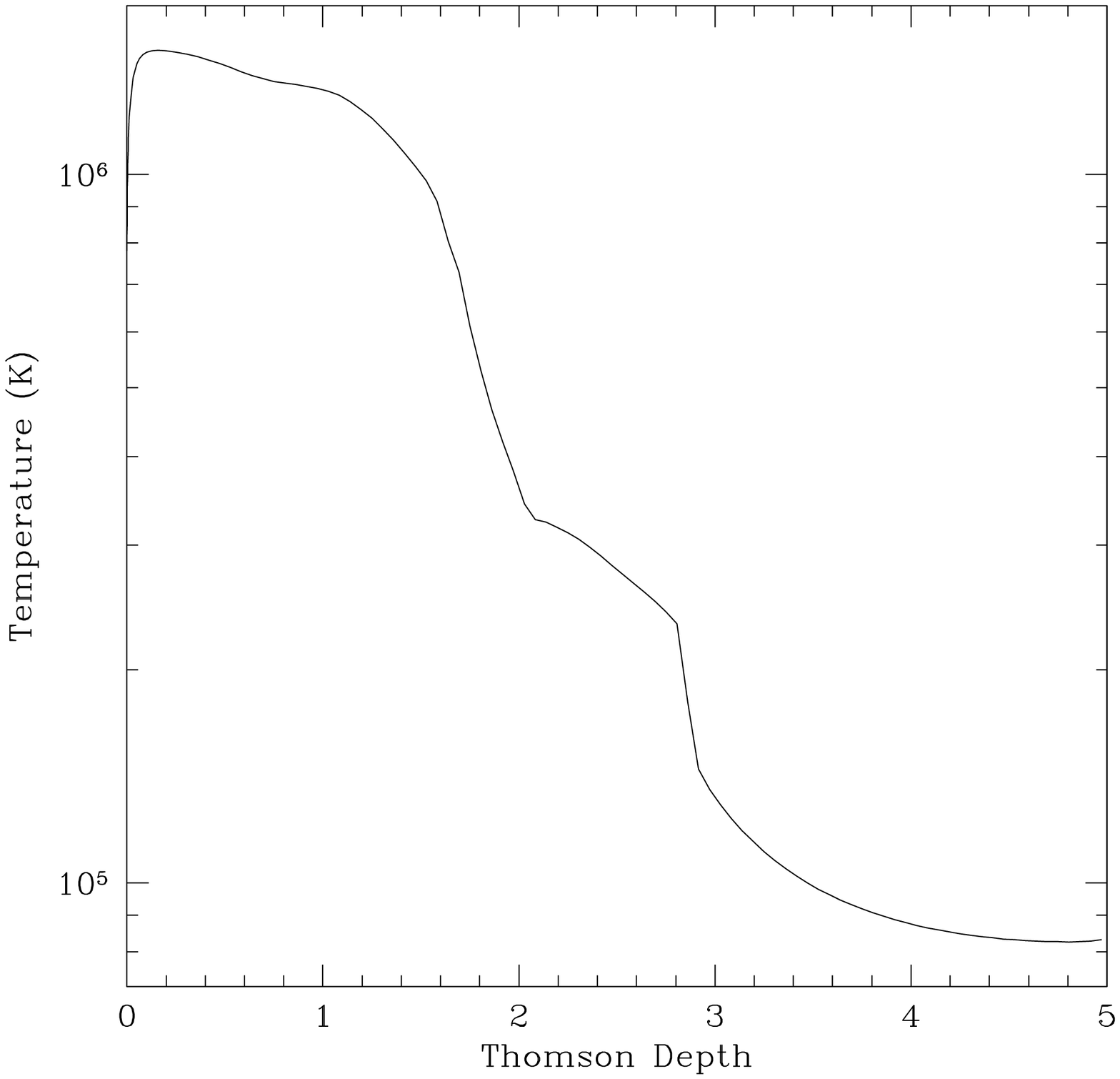,width=0.50\textwidth,silent=}}
\caption{{\it Top}: Reflection spectrum of a model with $\Gamma=2.5$ and
$F_x$ increased to $3 \times 10^{15}$~\ergcms. All the other parameters
are the same as the canonical model. The dashed line shows the illuminating
continuum, and the dotted line denotes the soft disc flux incident on the
bottom of the atmosphere. The reflection spectrum shows a strong He-like
Fe line at 6.7~\kev. {\it Bottom}: The temperature of the atmosphere as a
function of Thomson depth for the $\Gamma=2.5$ model. The temperature 
increase over the first 0.1 of a Thomson depth is due to photoelectric heating 
from the copious amounts of recombined Fe.}
\label{fig:gamma2.5}
\end{figure}

Narrow-Line Seyfert 1 Galaxies (NLS1s; Osterbrock \& Pogge 1985) are
an interesting sub-class of AGN with unusual X-ray properties (Brandt
1999; Leighly 1999a; Leighly 1999b; Vaughan \etal\ 1999b).  These
objects exhibit extreme X-ray variability, a strong soft excess, and
steep spectra in both the soft and hard X-ray bands. There are
indications from {\it ASCA} data that some of these object show
ionized features in their spectra (Turner, George \& Nandra 1998;
Vaughan \etal\ 1999a; Ballantyne \etal\ 2001). As discussed in
Sect.~\ref{sect:limits}, our code has difficulty computing the
structure of a weakly ionized atmosphere (such as one would get when
$\Gamma \ga 2.3$) without increasing the illuminating flux to
compensate for the lack of ionizing photons. In
Figure~\ref{fig:gamma2.5} we show the reflection spectrum and the
temperature structure for a canonical-like model with $\Gamma=2.5$ and
with the illuminating flux increased to $3 \times 10^{15}$~\ergcms.
The temperature structure of the model shows that the atmosphere is
relatively cool and barely heats up to just over a million degrees
(due to photoelectric heating of the recombined Fe ions), before
actually dropping to just under a 10$^6$~K at the surface. The
illuminating spectrum has little ionizing power, so that the Compton
temperature is low enough that the thermal instability is completely
skipped and the temperature falls smoothly with depth. As a result,
the reflection spectrum shows plenty of features from gas at a number
of different ionization states including a strong He-like Fe line at
6.7~\kev. These results show that our models predict ionized spectra
for steep continua and can be used for fitting the observations of
NLS1s.
  
\section{Conclusions}
\label{sect:concl}
Calculations of X-ray reflection spectra from ionized, optically-thick
material are an important tool for investigations of accretion flows
around compact objects. In this paper we have presented the results of
reflection calculations that have attempted to treat the relevant
physics with a minimum of assumptions.  We find that the transition
from hot, highly ionized gas to cold, neutral gas occurs over a small
but finite region in Thomson depth, and there is often a stable
temperature zone at $T \sim 2 \times 10^{6}$~K due to photoelectric
heating from recombining ions. As a consequence, our reflection spectra
often show features from partially ionized material, including
ionized \fe\ lines and edges. Then can be qualitatively described by
dilute versions of the older constant density reflection models.

Reflection models have great potential to constrain the properties of
accretion discs around luminous X-ray sources. The true test of our
models will come by fitting them to high signal-to-noise X-ray
spectral data of AGNs and GBHCs. Only by a comprehensive comparison
between models and data will real progress be made in understanding
the complex physics of accretion flows.

\section*{Acknowledgments}
We thank the referee, Chris Done, for useful and constructive comments.
DRB acknowledges financial support from the Commonwealth Scholarship
and Fellowship Plan and the Natural Sciences and Engineering Research
Council of Canada. RRR and ACF acknowledge support from the College of
the Holy Cross and the Royal Society, respectively.


\bsp 

\label{lastpage}

\end{document}